\documentclass[12pt,draftclsnofoot,onecolumn]{IEEEtran}
\usepackage[T1]{fontenc}
\usepackage[cmex10]{amsmath}
\usepackage[latin9]{inputenc}
\usepackage{xcolor}
\usepackage{pdfcolmk}
\usepackage{units}
\usepackage{color}
\usepackage{setspace}

\usepackage{amssymb}
\usepackage{epsfig}
\usepackage{stackrel}
\usepackage{graphicx}
\usepackage{esint}
\usepackage{cite}

\usepackage{epstopdf}
\usepackage{algorithm} 
\usepackage{algorithmic} 
\usepackage{textcomp} 

\begin{document}
\title{
{Multi-Path Cooperative Communications Networks for Augmented and Virtual Reality Transmission}
\vspace{0.1cm}}

\author{\normalsize
 Xiaohu Ge$^1$,~\IEEEmembership{Senior~Member,~IEEE,} Linghui Pan$^1$, Qiang Li$^1$,~\IEEEmembership{Member,~IEEE,} Guoqiang Mao$^2$,~\IEEEmembership{Senior~Member,~IEEE,} Song Tu$^1$\\
\vspace{0.70cm}
\small{
$^1$School of Electronic Information and Communications\\
Huazhong University of Science and Technology, Wuhan 430074, Hubei, P. R.
China.\\
Email: \{xhge, lhpan, qli\_patrick, songtu\}@mail.hust.edu.cn\\
\vspace{0.2cm}
$^2$School of Computing and Communications\\
The University of Technology Sydney, Australia.\\
Email: g.mao@ieee.org\\}

\thanks{\small{Submitted to IEEE transactions on multimedia.}}
\thanks{\small{Correspondence author: Dr. Qiang Li, Tel: +86 (0)27 87557942, Fax: +86 (0)27 87557943, Email: qli\_patrick@mail.hust.edu.cn.}}
}
\maketitle

\begin{abstract}

Augmented and/or virtual reality (AR/VR) are emerging as one of the main applications in future fifth generation (5G) networks. To meet the requirements of lower latency and massive data transmission in AR/VR applications, a solution with software-defined networking (SDN) architecture is proposed for 5G small cell networks. On this basis, a multi-path cooperative route (MCR) scheme is proposed to facilitate the AR/VR wireless transmissions in 5G small cell networks, in which the delay of MCR scheme is analytically studied. Furthermore, a service effective energy optimal (SEEO) algorithm is developed for AR/VR wireless transmission in 5G small cell networks. Simulation results indicate that both the delay and service effective energy (SEE) of the proposed MCR scheme outperform the delay and SEE of the conventional single path route scheme in 5G small cell networks.

\end{abstract}

\section{Introduction}
\label{sec1}

With the increasing popularity of various smart devices such as smart phones, tablets, Google Glass, Apple Watch, etc., augmented reality (AR) and virtual reality (VR) have the potential to be the next mainstream general computing platform \cite{R1}. Based on high-definition (HD) video services/applications, VR is able to mimic the real world by creating a virtual world or by applying various on-spot sensory equipments, such that immersive user experiences can be supported. In comparison, by augmenting more data information to the real world, AR is able to enhance the perception of reality in a manner of reality-plus-data. {\color{blue}Then, what is the real difference between VR and AR? For instance, with VR, you can dive and swim with dolphins. However, with AR, you can watch a dolphin pop out of your book. While VR provides an immersive environment for the user to interact with that world by a head-mounted display, AR makes the user see the superimposing content over the real world in mobile devices, such as laptops and smart phones.} Since real-time interactions and flows of massive information are involved in AR/VR applications, it will bring new challenges to the designs of future networks \cite{R2} for accommodating online AR/VR applications. To be specific, for future fifth generation (5G) wireless networks, the applications of AR/VR require innovations in the cloud-based network architectures \cite{R3}, with an objective of significantly improving the network throughput, delay performance, wireless capacity, etc.

To support the enormous traffic demands involved in AR/VR applications, there have been a great number of research activities in 5G network studies \cite{R4,R5,R6,R7}. N. Bhushan \emph{et al.} \cite{R4} pointed out that the 5G wireless networks could meet the 1000x traffic demands over the next decade, with additional spectrum availability, densification of small-cell deployments, and growth in backhaul infrastructures. Meanwhile, the massive multi-input multi-output (MIMO) technique offers huge advantages in terms of energy efficiency, spectral efficiency, robustness and reliability \cite{R5}, which allows for the use of low-cost hardware at both base stations and user terminals. Moreover, the large available bandwidth at millimeter wave (mmWave) frequencies makes mmWave transmission techniques attractive for the future 5G wireless networks \cite{R6,R7}. On the other hand, the gigantic data traffics on the next generation of Internet have also been investigated \cite{R8,R9,R10}. C. Walravens \emph{et al.} \cite{R8} studied the high-rate, small packet traffic in an Ethernet controller and used a technique called receive descriptor recycling (RDR) to reduce the small-packet loss by 40\%. N. Laoutaris \emph{et al.} \cite{R9} proposed using this already-paid-for off-peak capacity to perform global delay-tolerant bulk data transfers on the Internet. M. Villari \emph{et al.} studied the osmotic computing by moving cloud resources closer to the end users, which is regarded as a new paradigm for edge/cloud integration \cite{R10}. {\color{blue} What's more, some studies have been investigated for video communications \cite{N1,N2}, which play a significant role in big data communications. W. Xiang \emph{et al.} \cite{N1} proposed a light field (LF)-based 3D cloud telemedicine system and extend the standard multi-view video coding (MVC) to LF-MVC, which can achieve a significant 23 percent gain in compression ratio over the standard MVC approach. G. Wang \emph{et al.} \cite{N2} proposed a new LF-MVC prediction structure by extending the inter-view prediction into a two-directional parallel structure, which can achieve better coding performance and higher encoding efficiency.}

To reduce the network latency for high-speed reliable services like AR/VR and surveillance, usage of different chromatic dispersion compensation methods were discussed to reduce the transmission delay in fiber optical networks \cite{R11}. Low-delay rate control algorithms have been proposed to address the delay problem in \cite{R12,R13}. For instance, for the AR applications considered in \cite{R13}, low-delay communications of the encoded video over a Bluetooth wireless personal area network were investigated, by using a combination of the dynamic packetisation of video slices together with the centralized and predictive rate control. Mobile AR and VR have been used and studied recently \cite{R14,R15,R16,R17}. A. D. Hartl \emph{et al.} studied the verification of holograms by using mobile AR, where a re-parametrized user interface is proposed in \cite{R14}. The privacy preservation of cloth try-on was studied in \cite{R15}, by using mobile AR. In \cite{R18}, S. Choi \emph{et al.} proposed a prediction-based delay compensation system for head mounted display, which compensated delay up to 53 milliseconds with 1.083 degrees of minimum average error. In \cite{R19}, T. Langlotz \emph{et al.} presented an approach for mitigating the effect of color blending in optical see-through head-mounted displays, by introducing a real-time radiometric compensation.

Different from conventional video applications, AR/VR applications impose a strict requirement on the uplink/downlink transmission delays. Take the 360-degree video (an application of VR) as an example, the jitter and visual field delays, i.e., motion-to-photons (MTP) latency should be limited within 20 milliseconds. Otherwise, the users will feel dizzy \cite{R1}. Therefore, the transmission delay is one of main challenges for running AR/VR application over future 5G wireless networks. Moreover, few studies have investigated the issue of system energy consumption for the heavy traffic and high-rate transmissions required by AR/VR applications. Motivated by these gaps, it is important to propose a 5G wireless network solution for reducing the transmission delays and system energy consumption in AR/VR applications. The main contributions of this work are summarized in the following.
\begin{enumerate}
\item Based on requirements of AR/VR applications, a solution with software-defined networking (SDN) architecture is proposed for future 5G wireless networks, which is able to significantly reduce the network latency.
\item To facilitate the AR/VR data provisioning, a multi-path cooperative route (MCR) scheme is proposed for fast wireless transmissions from multiple edge data centers (EDCs) to the desired user. Moreover, the delay model with the MCR scheme is analytically studied. The lower and upper bounds of MCR delay are further obtained.
\item A service effective energy (SEE) model is proposed to evaluate the energy consumption of MCR scheme in AR/VR applications. Furthermore, a service effective energy optimization (SEEO) algorithm is developed for minimizing the SEE in 5G small cell networks.
\item Simulation results indicate that the delay and SEE of MCR scheme are better than the delay and SEE of conventional single path route scheme for AR/VR applications in 5G small cell networks.
\end{enumerate}

The remainder of the paper is organized as follows: In Section II, a solution with SDN architecture is proposed for AR/VR applications in 5G small cell networks, where storage strategies of AR/VR data are discussed. Section III describes a comprehensive network latency model. The MCR scheme is proposed and the MCR delay is investigated in Section IV. Moreover, the lower and upper bounds of MCR delay is derived. The optimization problem of SEE is formulated in Section V, where a two-step joint optimization algorithm is proposed for AR/VR applications in 5G small cell networks. Simulation results and discussions are presented in Section VI. Finally, Section VII concludes this paper.

\section{System Model}
\label{sec2}
\subsection{Network Model}
\begin{figure*}[!t]
\centerline{\includegraphics[width=17cm, draft=false]{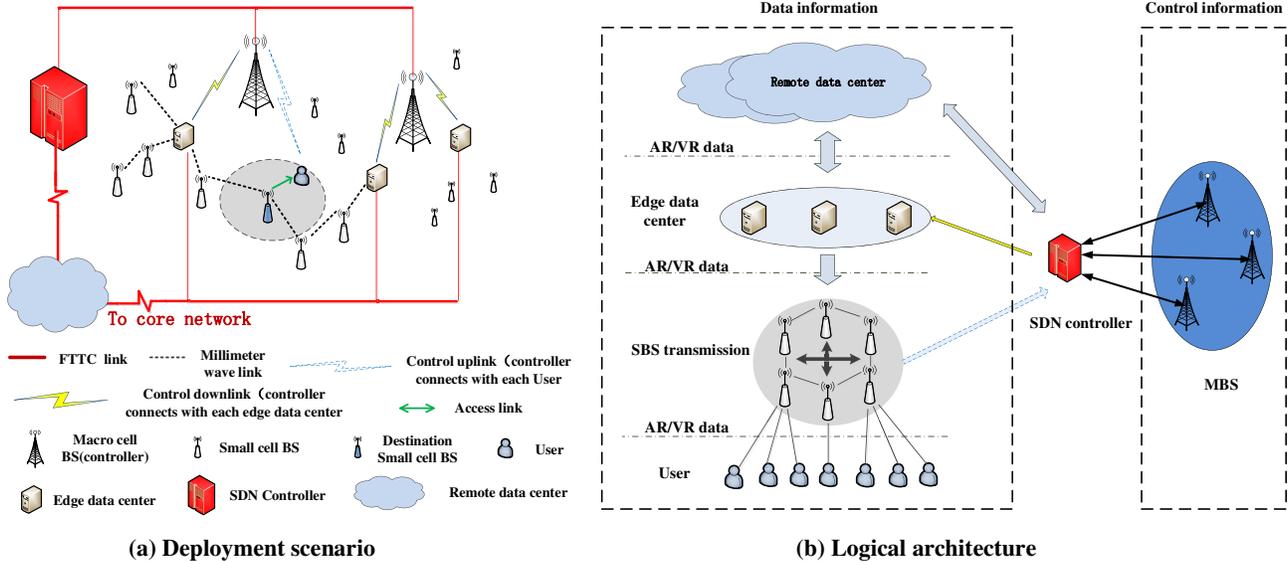}}
\caption{An SDN-based network architecture for 5G small cell networks.}
\label{fig1}
\end{figure*}

We consider a two-tier heterogeneous cellular network, where multiple small cell base stations (SBSs) and EDCs are deployed within the coverage of a macro cell base station (MBS). The MBS and EDCs are connected to the core networks through fiber to the cell (FTTC). An SDN architecture is adopted to support the separation of data and control information in Fig.~\ref{fig1}. As the network controller, the SDN controller divides the control information into MBSs and data information into SBSs and EDCs. MBSs are mainly responsible for the delivery of control information and routing decisions. The EDCs are data centers deployed at the edge of core networks, where massive AR/VR data are stored. All EDCs and SBSs, equipped with mmWave transmission techniques, are parts of the 5G small cell network. Utilizing the mmWave multi-hop transmission technique, AR/VR data streams can be transferred between EDCs and SBSs.

It is assumed that the MBSs, denoted by ${\Phi _M}$, follow a Poisson Point Process (PPP) with density ${\lambda _M}$, the SBSs, denoted by ${\Phi _S}$, follow a PPP with density ${\lambda _S}$, the user terminals, denoted by ${\Phi _U}$, follow a PPP with density ${\lambda _U}$, and the EDCs, denoted by ${\Phi _E}$, follow a PPP with density ${\lambda _E}$, respectively in the two-dimension (2-D) plane. The distributions of MBSs, SBSs, EDCs and user terminals are assumed to be independent of each other. To have a clear focus, only the coverage of a single MBS is analyzed in this paper. For ease of exposition, the SBSs located within the coverage of the MBS are denoted as a set ${\cal N}$, whose number is expressed as {\color{blue} $\left| {\cal N} \right|$} . The set of EDCs located within the coverage of the MBS is denoted as ${\cal M}$, whose number is expressed as {\color{blue} $\left| {\cal M} \right|$}. The set of user terminals within the coverage of the MBS is denoted as ${\cal U}$, whose number is expressed as $\left| {\cal U} \right|$.

For a user terminal with AR/VR traffic demands, a request is first sent to an associated MBS by uplinks. Upon receiving a user's request, the associated MBS searches EDCs that are located close to the requesting user. If however, there is an EDC located outside the macro cell, then the associated MBS sends the request to the MBS to which this EDC belongs, through the SDN controller. Upon receiving the routing information transmitted from the MBS, the close-by EDCs transmit AR/VR data to the destination SBS that is located closest to the requesting user. Finally, the destination SBS delivers the AR/VR data to the requesting user by mmWave transmission links. In above solution, the SDN architecture is adopted to facilitate the request/feedback transmissions from users by control information links. Moreover, the big AR/VR data is promptly transmitted by multiple EDCs to the destination SBS. Utilizing the buffering scheme of the destination SBS, the jitter of multi-path AR/VR transmission between EDCs and the destination is minimized. Furthermore, the requirements of AR/VR applications, e.g. low network latency and massive data transmission, are satisfied by this SDN solution in 5G small cell networks.

\subsection{Storage Strategies of AR/VR Data}
Consider a library ${\cal K}$ that consists of $\left| {\cal K} \right|$ AR/VR video contents. We define ${\mathbf{q}} = \left[ {{q_1},{q_2},...,{q_{\left| {\cal K} \right|}}} \right]$ as the corresponding popularity distributions of these $\left| {\cal K} \right|$ AR/VR video contents that are arranged in a popularity descending order, where ${q_k}$ ($k \in {\cal K}$ and $1 \leqslant k \leqslant \left| {\cal K} \right|$) denotes the popularity of the $k - th$ popular AR/VR video content, i.e., the probability that an arbitrary user request is for AR/VR video content $k$. Without loss of generality, the distribution of ${q_k}$ is assumed to be governed by a Zipf distribution \cite{R21},

\[{q_k} = \frac{{{k^{ - \beta }}}}{{\sum\nolimits_{j = 1}^{\left| {\cal K} \right|} {{j^{ - \beta }}} }},\tag{1}\]
where $\beta $ is a positive value that characterizes the skewness of the popularity distribution, and $\sum\limits_{k = 1}^{\left| {\cal K} \right|} {{q_k} = 1} $.

In general, a subset of the $\left| {\cal K} \right|$ AR/VR video contents are stored in EDCs, whereas the remaining video contents can be fetched from the remote data centers (RDC). For ease of exposition, we let ${\mathbf{x}} = \left( {{x_{mk}} \in \{ 0,1\} :ED{C_m} \in {\cal M},k \in {\cal K}} \right)$ denote the AR/VR video content placement strategy in EDCs, where ${x_{mk}} = 1$ means that AR/VR video content  $k$ ($k \in {\cal K}$) is stored at EDC  $ED{C_m}$ ($ED{C_m} \in {\cal M}$), whereas ${x_{mk}} = 0$ means that AR/VR video content $k$ is not stored at EDC $ED{C_m}$. Since the storage capacity at each EDC is limited that cannot store all $\left| {\cal K} \right|$ AR/VR video contents, generally there exist three storage strategies in existing studies.
\begin{enumerate}
\item First in first out (FIFO): The contents stored at the EDC form a queue in a chronological order until the storage capacity is reached. In this way, the content at the head of the queue has to be removed for accommodating the newly incoming content at the tail of the queue.
\item Least recently used (LRU): The contents are stored at the EDC according to how frequently they are requested by users recently. In this way, if the storage capacity of EDC is reached, then the least recently used content will be replaced by the newly incoming content.
\item Popularity-priority: All contents are stored at the EDC in a popularity descending order until the storage capacity is reached. In this way, the more popular content can be stored by comparing the popularity of the newly incoming content with that of the least popular content stored at EDC.
\end{enumerate}

In our considered network model, the popularity-priority strategy is adopted to store the most popular AR/VR video content data at each EDC. Although the popularity distribution of AR/VR video contents may vary with time, to maintain cache consistency, we consider in this paper a relatively static popularity distribution within a given interval. This assumption is valid in examples including popular short news videos that are updated every 2-3 hours, new movies that are posted every week, new music videos that are posted about every month.

For ease of analysis, it is assumed that each EDC has the same storage capacity, in which $\Psi$ ($\Psi \leqslant \left| {\cal K} \right|$) AR/VR video contents of the same size can be stored. Since the popularity-priority strategy is adopted where each EDC stores the most popular contents independently, it ends up with that the same $\Psi$ most popular video contents are stored at each EDC, i.e.,
\[\sum\limits_{k = 1}^\Psi {{x_{mk}}}  = \Psi,\forall ED{C_m} \in {\cal M}.\tag{2}\]

For ease of reference, the notations and symbols with the default values used in this paper and simulations are listed in Table~\ref{tab1}.
\begin{table*}
\centering
\caption{Notations and Symbols with Default Values}
\scalebox{0.8}{
\begin{tabular}{l|l|l}
\hline Symbols & Definition/explanation & Default values \\
\hline
${\lambda _M}$, ${\lambda _S}$, ${\lambda _U}$ & Density of MBSs, SBSs, users, respectively & 5 ${\text{k}}{{\text{m}}^{-2}}$, 50 ${\text{k}}{{\text{m}}^{-2}}$, 200 ${\text{k}}{{\text{m}}^{-2}}$  \\
\hline
${P_S}$, ${P_E}$, ${P_U}$ & Transmission power at each SBS, EDC, user, respectively & 30 dBm, 30 dBm, 23 dBm  \\
\hline
${\tau _{mmW}}$ & Time slot of millimeter wave channels & 5 $\mu$s  \\
\hline
${N_0}$ & Noise power spectral density & -174 ${\text{dBm}} \cdot {\text{H}}{{\text{z}}^{ - 1}}$  \\
\hline
${W_{mmW}}$ & Bandwidth of millimeter wave wireless links & 200 MHz  \\
\hline
$\mu$ & Service rate at the MBS & $1.05 \times {10^4}{\kern 1pt} {\kern 1pt}  {\kern 1pt} {{\text{s}}^{ - 1}}$  \\
\hline
$\chi$ & Factor of user arrival rate & $5 \times {10^7}{\kern 1pt} {\kern 1pt} {{\text{m}}^{\text{2}}} \cdot {{\text{s}}^{ - 1}}$  \\
\hline
$\sigma$ & Standard variance of shadow fading over wireless channels & 5 dB  \\
\hline
${r_{mmW}}$ & Maximum transmission distance of SBSs and EDCs & 100 meters  \\
\hline
$L$ & Data packet size & 1024 Bytes  \\
\hline
$\Omega$ & Buffer size of SBS & 1 MB  \\
\hline
${L_{fiber}}$ & Distance between the macro cell network and the RDC & 1000 km  \\
\hline
${v_{fiber}}$ & Transmission rate in optic fibers & $2 \times {10^8}{\kern 1pt} {\kern 1pt} {\kern 1pt} {\text{m}} \cdot {{\text{s}}^{ - 1}}$  \\
\hline
$\beta$ & Skewness parameter of popularity distributions & 0.8  \\
\hline
$\left| {\cal K} \right|$ & Number of video contents in the library ${\cal K}$ & 500  \\
\hline
${R_{\max }}$ & Maximum distance between the EDC and the destination SBS & 500 meters  \\
\hline
${a_M}$, ${b_M}$ & The fixed coefficient of MBS & 21.45, 354 Watt  \\
\hline
${a_S}$, ${b_S}$ & The fixed coefficient of SBS & 7.84, 71 Watt  \\
\hline
${a_E}$, ${b_E}$ & The fixed coefficient of EDC & 7.84, 71 Watt  \\
\hline
$T_{Lifetime}^M$ & Lifetime of MBS & 10 years  \\
\hline
$T_{Lifetime}^S$ & Lifetime of SBS & 5 years  \\
\hline
$T_{Lifetime}^E$ & Lifetime of EDC & 5 years  \\
\hline
${E_{storage}}$ & Energy consumption of one video content stored at the EDC & $8 \times {\text{1}}{{\text{0}}^6}{\kern 1pt} $ Joule  \\
\hline
\end{tabular}
}
\label{tab1}
\end{table*}

\section{Network Latency Model}
For the AR/VR service provisioning upon receiving an arbitrary user's request, the network latency model is formulated as
{\small{
\[\begin{gathered}
\begin{aligned}
  D &= \left( {D_{UL}^{req} + D_{DL}^{deli} + D_{DL}^{bh} + D_{DL}^{as}} \right) \cdot {P_{in - EDC}} + \left( {D_{UL}^{req} + D_{DL}^{deli} + D_{DL}^{bh} + D_{DL}^{as} + {D^{fiber}}} \right) \cdot \left( {1 - {P_{in - EDC}}} \right) \hfill \\
  &= D_{UL}^{req} + D_{DL}^{deli} + D_{DL}^{bh} + D_{DL}^{as} + {D^{fiber}} \cdot \left( {1 - {P_{in - EDC}}} \right) \hfill \\
\end{aligned}
\end{gathered},\normalsize{\tag{3}}\]}}\\
with
\[{P_{in - EDC}} = \sum\limits_{k = 1}^\Psi {{q_k}},\tag{4}\]
where ${P_{in - EDC}}$ denotes the probability that the requesting user can find the desired AR/VR data in the nearby EDCs. If the required AR/VR data can be found in EDCs, then the network latency is a sum of the delays within a macro cell coverage. Otherwise, the transmission delay over the optical networks should also be taken into account.

(1) $D_{UL}^{req}$: The delay incurred when a user terminal sends a request to the MBS through uplinks, which consists of the uplink transmission delay $D_{UL}^{req,1}$ and the queuing delay at the MBS $D_{UL}^{req,2}$, i.e.,
\[D_{UL}^{req} = D_{UL}^{req,1} + D_{UL}^{req,2}.\tag{5}\]

When a user $u \in {\cal U}$ sends a request to the MBS, the corresponding received signal power at the MBS is given as \cite{R26}
\[\gamma _{UL}^{req,1} = {P_U}r_o^{ - {\alpha _1}}{\left| {{{\mathbf{H}}_{uo}}} \right|^2},\tag{6}\]
where the same transmission power ${P_U}$ is adopted at each user $u \in {\cal U}$, ${r_o}$ denotes the distance between the user $u$ and MBS $MB{S_o}$, ${\alpha _1}$ denotes the path-loss exponent, ${{\mathbf{H}}_{uo}} \in {{{\mathbb{C}}}^{n_r^M \times n_t^U}}$ denotes the channel gains from the user $u$ to the MBS $MB{S_o}$, where we have ${h_{uo,{x_1},{y_1}}} \sim {{{\cal C}{\cal N}}}(0,1)$ for each entry ${h_{uo,{x_1},{y_1}}}$ ($ {x_1} = 1,2,...,n_r^M$, ${y_1} = 1,2,...,n_t^U$), and $n_r^M$  and $n_t^U$ denote the number of receive antennas at the MBS and the number of transmission antennas at the user, respectively. Then the probability of successfully delivering the user request to the MBS is expressed as
\[\begin{gathered}
\begin{aligned}
  \rho _{UL}^{req,1} &= \Pr (\gamma _{UL}^{req,1} \geqslant {\theta _1}) \hfill \\
  &= \Pr ({P_U}r_o^{ - {\alpha _1}}{\left| {{{\mathbf{H}}_{uo}}} \right|^2} \geqslant {\theta _1}) \hfill \\
  &\mathop  = \limits^{(a)} \Pr (\frac{{{P_U}r_o^{ - {\alpha _1}}(\sum\limits_{{x_1} = 1}^{n_r^M} {\sum\limits_{{y_1} = 1}^{n_t^U} {{{\left| {{h_{uo,{x_1},{y_1}}}} \right|}^2}} } )}}{{n_t^U}} \geqslant {\theta _1}) \hfill \\
  &= \Pr (\frac{{{P_U}r_o^{ - {\alpha _1}}g}}{{n_t^U}} \geqslant {\theta _1}) \hfill \\
  &= \Pr (g \geqslant \frac{{{\theta _1}n_t^Ur_o^{{\alpha _1}}}}{{{P_U}}}) \hfill \\
  &= {{{\mathbb{E}}}_{{r_o}}}\left( {\sum\limits_{t = 0}^{n_t^Un_r^M - 1} {\frac{1}{{t!}}{{\left( {\frac{{{\theta _1}n_t^Ur_o^{{\alpha _1}}}}{{{P_U}}}} \right)}^t}{e^{ - \frac{{{\theta _1}n_t^Ur_o^{{\alpha _1}}}}{{{P_U}}}}}} } \right) \hfill \\
  &= \int_0^\infty  {\left( {\sum\limits_{t = 0}^{n_t^Un_r^M - 1} {\frac{{2\pi {\lambda _M}r}}{{t!}}{{\left( {\frac{{{\theta _1}n_t^U{r^{{\alpha _1}}}}}{{{P_U}}}} \right)}^t}{e^{ - \left( {\frac{{{\theta _1}n_t^U{r^{{\alpha _1}}}}}{{{P_U}}} + {\lambda _M}\pi {r^2}} \right)}}} } \right)} dr \hfill \\
\end{aligned}
\end{gathered},\tag{7}\]
where ${\theta _1}$ denotes the received signal power threshold for successful reception at the MBS. Provided that MBSs are randomly located in a 2-D plane with density ${\lambda _M}$, approximation $(a)$ is obtained based on [Eq. (10), 22], where we have $g \sim Gamma\left( {n_t^Un_r^M,1} \right)$ by letting $g = \sum\limits_{{x_1} = 1}^{n_r^M} {\sum\limits_{{y_1} = 1}^{n_t^U} {{{\left| {{h_{uo,{x_1},{y_1}}}} \right|}^2}} } $. Otherwise if $\gamma _{UL}^{req,1} < {\theta _1}$, retransmissions are performed. Then we have the average transmission delay for successfully delivering a user request to the MBS as
\[D_{UL}^{req,1} = \frac{{T_{UL}^{req,1}}}{{\rho _{UL}^{req,1}}},\tag{8}\]
where $T_{UL}^{req,1}$ denotes the time for transmitting a user request from a user to the associated MBS, and $\frac{1}{{\rho _{UL}^{req,1}}}$ corresponds to the average number of retransmissions \cite{R23}.

Furthermore, an M/M/1 queuing model is adopted for calculating the queuing delay $D_{UL}^{req,2}$ at the MBS upon successfully receiving the user's request. Then we have
\[D_{UL}^{req,2} = \frac{1}{{\mu  - {\lambda _a}}},\tag{9}\]
where ${\lambda _a}$ denotes the average arrival rate of users' requests at the MBS, which is proportional to the user density in the macro cell, i.e., ${\lambda _a} = \chi {\lambda _U}$ where $\chi $ is the factor of user arrival rate. $\mu $ denotes the service rate at the MBS.

(2) $D_{DL}^{deli}$: The delay incurred when the MBS sends the user request/control information to the EDCs that own the requested AR/VR data through downlinks.

When the MBS sends routing information to the specifical EDC $EDC_m^u$ (${\text{ }}EDC_m^u \in {\cal M}$) that owns the AR/VR data requested by user $u \in {\cal U}$, the corresponding received signal-to-interference-and-noise ratio (SINR) at EDC is given by
\[\gamma _{DL}^{deli} = \frac{{{P_M}r_o^{ - {\alpha _1}}{{\left| {{{\mathbf{H}}_{ou}}} \right|}^2}}}{{\sum\limits_{MB{S_l} \in {\Phi _M},MB{S_l} \ne MB{S_o}} {{P_M}r_l^{ - {\alpha _1}}{{\left| {{{\mathbf{H}}_{lu}}} \right|}^2} + \sigma _z^2} }},\tag{10}\]
where ${P_M}$ denotes the transmission power at MBS, ${r_o}$ denotes the distance between MBS $MB{S_o}$ and EDC $EDC_m^u$, ${r_l}$ denotes the distance between MBS $MB{S_l}$ and EDC $EDC_m^u$, $\sigma _z^2$ denotes the noise power, ${{\mathbf{H}}_{ou}} \in {{{\mathbb{C}}}^{n_r^E \times n_t^M}}$ denotes the channel gains between MBS $MB{S_o}$ and EDC $EDC_m^u$, where we have ${h_{ou,{x_2},{y_2}}} \sim {\cal C}{\cal N}(0,1)$ for each entry ${h_{ou,{x_2},{y_2}}}$(${x_2} = 1,2,...,n_r^E$, ${y_2} = 1,2,...,n_t^M$), ${{\mathbf{H}}_{lu}} \in {{{\mathbb{C}}}^{n_r^E \times n_t^M}}$ denotes the channel gains between MBS $MB{S_l}$ and EDC $EDC_m^u$, where we have ${h_{lu,{x_2},{y_2}}} \sim {\cal C}{\cal N}(0,1)$ for each entry ${h_{lu,{x_2},{y_2}}}$(${x_2} = 1,2,...,n_r^E$, ${y_2} = 1,2,...,n_t^M$), and $n_r^E$ and $n_t^M$ denote the number of receive antennas at the EDC and the number of transmission antennas at the MBS, respectively. Based on our previous work \cite{R22}, the probability of successfully delivering the routing information from the MBS to EDC $EDC_m^u$ can be readily expressed as
{\footnotesize{
\[\begin{gathered}
\begin{aligned}
  \rho _{DL}^{deli} &= \Pr (\gamma _{DL}^{deli} \geqslant {\theta _2}) \hfill \\
  &= \Pr (\frac{{{P_M}r_o^{ - {\alpha _1}}{{\left| {{{\mathbf{H}}_{ou}}} \right|}^2}}}{{\sum\limits_{MB{S_l} \in {\Phi _M},MB{S_l} \ne MB{S_o}} {{P_M}r_l^{ - {\alpha _1}}{{\left| {{{\mathbf{H}}_{lu}}} \right|}^2} + \sigma _z^2} }} \geqslant {\theta _2}) \hfill \\
  &= \int_0^\infty  {2\pi {\lambda _M}r{e^{ - \pi {\lambda _M}{r^2}}}\left( {\exp \left( { - \pi {\lambda _M}{r^2}\theta _2^{\frac{2}{{{\alpha _1}}}}\int_{\theta _2^{ - \frac{2}{{{\alpha _1}}}}}^\infty  {\left( {1 - \frac{1}{{{{\left( {1 + {v^{ - \frac{{{\alpha _1}}}{2}}}} \right)}^{n_t^Mn_r^E}}}}} \right)dv} } \right) + sum\left( {{{\mathbf{x}}_{n_t^Mn_r^E - 1}}} \right)} \right)dr}  \hfill \\
\end{aligned}
\end{gathered},\normalsize{\tag{11}}\]
}}\\
where ${\theta _2}$ denotes the SINR threshold for successful reception at the EDC and $sum\left( {{{\mathbf{x}}_{n_t^Mn_r^E - 1}}} \right)$ denotes the summation of all elements in ${{\mathbf{x}}_{n_t^Mn_r^E - 1}} \in {{{\mathbb{C}}}^{(n_t^Mn_r^E - 1) \times 1}}$. For ease of exposition, we let

\[{{\mathbf{x}}_{n_t^Mn_r^E}} = {\left[ {{x_1},{x_2},...,{x_{n_t^Mn_r^E}}} \right]^T},\tag{12}\]

\[\begin{gathered}
\begin{aligned}
  {{\mathbf{y}}_{n_t^Mn_r^E}} &= {\left[ {{y_1},{y_2},...,{y_{n_t^Mn_r^E}}} \right]^T} \hfill \\
  &= {\left[ {n_t^Mn_r^E{k_1},\frac{{n_t^Mn_r^E\left( {1 + n_t^Mn_r^E} \right)}}{2}{k_2},...,C_{2n_t^Mn_r^E - 1}^{n_t^Mn_r^E} \cdot {k_{n_t^Mn_r^E}}} \right]^T} \hfill \\
\end{aligned}
\end{gathered},\tag{13}\]

\[{{\mathbf{G}}_{n_t^Mn_r^E}} = \left[ {\begin{array}{*{20}{c}}
  0&{}&{}&{}&{} \\
  {\frac{1}{2}n_t^Mn_r^E{k_1}}&0&{}&{}&{} \\
  {\frac{1}{3}n_t^Mn_r^E\left( {1 + n_t^Mn_r^E} \right){k_2}}&{\frac{1}{3}n_t^Mn_r^E{k_1}}&0&{}&{} \\
   \vdots &{}&{}&0&{} \\
  {\frac{{n_t^Mn_r^EC_{2n_t^Mn_r^E - 2}^{n_t^Mn_r^E}}}{{n_t^Mn_r^E}}{k_{n_t^Mn_r^E - 1}}}&{\frac{{n_t^Mn_r^EC_{2n_t^Mn_r^E - 3}^{n_t^Mn_r^E}}}{{n_t^Mn_r^E}}{k_{n_t^Mn_r^E - 2}}}& \cdots &{\frac{{n_t^Mn_r^EC_{n_t^Mn_r^E}^{n_t^Mn_r^E}}}{{n_t^Mn_r^E}}{k_1}}&0
\end{array}} \right].\tag{14}\]
Then ${{\mathbf{x}}_{n_t^Mn_r^E}}$ can be expressed by using (13) and (14) as
\[{{\mathbf{x}}_{n_t^Mn_r^E}} = \sum\limits_{t = 1}^{n_t^Mn_r^E} {{{\left( {\pi {\lambda _M}{r^2}} \right)}^t}} {x_0}{\mathbf{G}}_{_{n_t^Mn_r^E}}^{t - 1}{{\mathbf{y}}_{n_t^Mn_r^E}},\tag{15a}\]
and
\[{{\mathbf{x}}_{n_t^Mn_r^E - 1}} = \sum\limits_{t = 1}^{n_t^Mn_r^E - 1} {{{\left( {\pi {\lambda _M}{r^2}} \right)}^t}} {x_0}{\mathbf{G}}_{_{n_t^Mn_r^E - 1}}^{t - 1}{{\mathbf{y}}_{n_t^Mn_r^E - 1}},\tag{15b}\]
where
\[{x_0} = {e^{ - \pi {\lambda _M}{k_0}{r^2}}},\tag{15c}\]

\[{k_0} = \theta _2^{\frac{2}{{{\alpha _1}}}}\int_{\theta _2^{ - \frac{2}{{{\alpha _1}}}}}^\infty  {\left( {1 - \frac{1}{{{{\left( {1 + {v^{ - \frac{{{\alpha _1}}}{2}}}} \right)}^{n_t^Mn_r^E}}}}} \right)} dv,\tag{15d}\]

\[{k_q} = \theta _2^{\frac{2}{{{\alpha _1}}}}\int_{\theta _2^{ - \frac{2}{{{\alpha _1}}}}}^\infty  {\frac{1}{{{{\left( {1 + {v^{\frac{{{\alpha _1}}}{2}}}} \right)}^q}{{\left( {1 + {v^{ - {\kern 1pt} \frac{{{\alpha _1}}}{2}}}} \right)}^{n_t^Mn_r^E}}}}} dv,q \geqslant 1,\tag{15e}\]
respectively. Then we have
\[D_{DL}^{deli} = \frac{{T_{DL}^{deli}}}{{\rho _{DL}^{deli}}},\tag{16}\]
where $T_{DL}^{deli}$ denotes the downlink transmission time of a packet from the MBS to the EDC.

(3) $D_{DL}^{bh}$: The backhaul delay, i.e., the delay incurred when the EDC delivers the requested AR/VR data to the destination SBS which associates with the requesting user.

In order to reduce the transmission delay of AR/VR data in 5G small cell networks, a multi-path cooperative route (MCR) scheme is proposed in Section IV. A, where the closest EDCs simultaneously transmit the massive AR/VR data to the destination SBS.

(4) $D_{DL}^{as}$: The delay incurred when the destination SBS transmits the requested AR/VR data to the requesting user. Considering the directivity of mmWave transmission, the interference is negligible in 5G small cell networks \cite{R24}. Then the corresponding received signal power at user $u$ is given by
\[\gamma _{DL}^{as} = {P_S}r_u^{ - {\alpha _2}}{\left| {{{\mathbf{H}}_{nu}}} \right|^2},\tag{17}\]
where the same transmission power ${P_S}$ is adopted at each SBS, ${r_u}$ denotes the distance between the destination SBS $SBS_n^u\left( {SBS_n^u \in {\cal N}} \right)$ and the user $u$, ${\alpha _2}$ denotes the path-loss exponent, ${{\mathbf{H}}_{nu}} \in {{{\mathbb{C}}}^{n_r^U \times n_t^S}}$ denotes the channel gains between the SBS $SBS_n^u$ and the user $u$, where we have ${h_{nu,{x_3},{y_3}}} \sim {\cal C}{\cal N}(0,1)$ for each entry ${h_{nu,{x_3},{y_3}}}$(${x_3} = 1,2,...,n_r^U$, ${y_3} = 1,2,...,n_t^S$), and $n_r^U$ and $n_t^S$ denote the number of receive antennas at the user and the number of transmission antennas at the SBS, respectively. The probability of successfully delivering the requested AR/VR data to the requesting user is expressed as
\[\begin{gathered}
\begin{aligned}
  \rho _{DL}^{as} &= \Pr (\gamma _{DL}^{as} \geqslant {\theta _3}) \hfill \\
  &= \Pr ({P_S}r_u^{ - {\alpha _2}}{\left| {{{\mathbf{H}}_{nu}}} \right|^2} \geqslant {\theta _3}) \hfill \\
  &= \int_0^\infty  {\left( {\sum\limits_{t = 0}^{n_t^Sn_r^U - 1} {\frac{{2\pi {\lambda _S}r}}{{t!}}{{\left( {\frac{{{\theta _3}n_t^S{r^{{\alpha _2}}}}}{{{P_S}}}} \right)}^t}{e^{ - \left( {\frac{{{\theta _3}n_t^S{r^{{\alpha _2}}}}}{{{P_S}}} + {\lambda _S}\pi {r^2}} \right)}}} } \right)} dr \hfill \\
\end{aligned}
\end{gathered},\tag{18}\]
where ${\theta _3}$ denotes the received signal power threshold for a successful reception. Then we have

\[D_{DL}^{as} = \frac{{T_{DL}^{as}}}{{\rho _{DL}^{as}}},\tag{19}\]
where $T_{DL}^{as}$ denotes the downlink transmission time of a data packet from the destination SBS to the user.

(5) ${D^{fiber}}$: The fiber delay incurred in the uplink/downlink transmissions when the MBS has to fetch AR/VR data from the RDC upon a search failure in local EDCs. Considering that the transmission rate in optic fibers is ${v_{fiber}}$ and the distance between the given macro cell and the RDC is ${L_{fiber}}$, the fiber delay is expressed as
\[{D^{fiber}} = \frac{{2{L_{fiber}}}}{{{v_{fiber}}}}.\tag{20}\]

\section{AR/VR Multi-path Cooperative Transmissions}
For AR/VR applications, not only a large network throughput is required for transmitting a massive data, but also a low system delay is needed to support the user interactions. In traditional networks the data is transmitted from a source to a destination through a fixed path. In this case, the maximum network throughput is restricted by the minimum transmission rate of links along the path. It implies that the total network throughput is constrained by the most congested link with the minimum transmission rate in the fixed path. Moreover, the system delay is also limited by the minimum transmission rate in the bottleneck of the fixed path. To solve these issues, a MCR scheme is proposed to meet the requirements of the massive data transmission and low system delay for AR/VR applications.
\subsection{Multi-Path Cooperative Route (MCR) Scheme}
Considering the fluctuation of wireless channels, it is very difficult to transmit the massive AR/VR data with the low system delay constraint by a fixed path in 5G small cell networks. On the other hand, the AR/VR data can be repeatedly stored in multiple EDCs according to the content popularity. Therefore, the same AR/VR data can be cooperatively transmitted to a user from adjacent EDCs. The basic multi-path cooperative route scheme is described as follows:
\begin{enumerate}
\item EDCs selection: According to the system model, multiple EDCs are located in a macro cell. When the requested AR/VR data is stored at EDCs, $B$ selected EDCs simultaneously transmit the same AR/VR data to a destination SBS that is the closest to the requesting user. In the end, this destination SBS transmits the AR/VR data to the user by mmWave links.
\item Multi-path transmission strategy: $B$ selected EDCs are incrementally ordered by the average distance ${r_i},{\text{ }}1 \leqslant i \leqslant B$ between the EDC $ED{C_i}$ and the destination SBS. The $ED{C_p},{\text{ }}1 \leqslant p \leqslant B,$ is the selected EDC away from the destination SBS with an average distance ${r_p}$. The requested AR/VR data $\mathfrak{S}$ is divided into $B$ parts with the proportion $\frac{{\frac{1}{{{r_p}}}}}{{\sum\limits_{i = 1}^B {\frac{1}{{{r_i}}}} }}$. Moreover, the selected EDC $ED{C_p}$ only need to transmit the data of $\frac{{\frac{1}{{{r_p}}}}}{{\sum\limits_{i = 1}^B {\frac{1}{{{r_i}}}} }} \cdot \mathfrak{S}$. In this case, the closer EDC need to transmit the larger AR/VR data and the distant EDC transmit a smaller amount of AR/VR data.
\item Relay SBSs selection: Based on the system model, SBSs are densely deployed in the coverage of every MBS. When the maximum transmission distance of SBS with mmWave transmission techniques is configured as ${r_{mmW}}$, it is assumed that there exist more than two SBSs in the distance ${r_{mmW}}$. In this case, the requested AR/VR data is transmitted to the destination SBS by wireless relayed SBSs. To minimize the relay delay in SBSs, the relay route algorithm with the minimum hop number is adopted for the transmission path between the requested EDC and the destination SBS. Relay SBSs are selected by the relay routing algorithm with the minimum hop number, e.g., the shortest path based geographical routing algorithm \cite{R27}.
\end{enumerate}

\subsection{Delay Theorem of Multi-Path Cooperative Route Scheme}
\textbf{MCR Delay Theorem}: When EDCs are deployed in a 5G small cell network, the AR/VR data stored at $B$ EDCs are simultaneously transmitted to a destination SBS by the MCR scheme. The system delay of MCR scheme is expressed by
\[D_{DL}^{bh} = \frac{{\left\lceil {\frac{\Omega}{L}} \right\rceil }}{{\sum\limits_{i = 1}^B {\frac{1}{{{r_i}}}} }} \cdot \frac{{2{\tau _{mmW}}\left( {1 + 1.28\frac{{{\lambda _S}}}{{{\lambda _E}}}} \right)}}{{{r_{mmW}}\left( {1 + erf\left( {\frac{{f\left( {{r_{mmW}}} \right)}}{{\sqrt 2 \sigma }}} \right)} \right)}},\tag{21a}\]
\[{r_i} = \frac{{\left( {2i - 1} \right) \cdot \left( {2i - 3} \right) \cdots 1}}{{{2^i}\sqrt {{\lambda _e}}  \cdot \Gamma \left( i \right)}},\tag{21b}\]
\[f\left( {{r_{mmW}}} \right)\left( {{\text{dB}}} \right) = {P_S}\left( {{\text{dB}}} \right) - {\theta _4}\left( {{\text{dB}}} \right) - {N_0}{W_{mmW}}\left( {{\text{dB}}} \right) - 70 - 20{\log _{10}}\left( {{r_{mmW}}} \right),\tag{21c}\]
where $\Omega$ is the buffer size of SBS, $L$ is the data packet size, $\left\lceil  \cdot  \right\rceil $ is the rounding up operation of a number, ${r_i}$ is the average distance between the destination SBS and the EDC $ED{C_i}$. ${\lambda _S}$ and ${\lambda _E}$ are the densities of SBSs and EDCs in the coverage of MBS. In a wireless route between the EDC and the destination SBS, the wireless transmission is time slotted and one packet is transmitted in each time slot ${\tau _{mmW}}$. ${r_{mmW}}$ is the maximum transmission distance of SBSs and EDCs, ${P_S}\left( {{\text{dB}}} \right)$ is the transmission power of a SBS, ${\theta _4}\left( {{\text{dB}}} \right)$ is the threshold of the signal receiver, ${N_0}$ is the noise power spectral density, ${W_{mmW}}$ is the bandwidth of mmWave links, $\sigma$ is the standard deviation of shadow fading over wireless channels in dB.

\emph{\textbf{Proof:}} According to the MCR scheme, the destination SBS simultaneously receives the AR/VR data from adjacent EDCs by multi-path wireless routes. ${R_p}$ is the distance between the destination SBS and the EDC $ED{C_p}$. Without loss of generality, the probability density function (PDF) of the distance ${R_p}$ is assumed to be governed by \cite{R22}
\[{f_{{R_p}}}\left( r \right) = {e^{ - {\lambda _E}\pi {r^2}}}\frac{{2{{\left( {{\lambda _E}\pi {r^2}} \right)}^p}}}{{r\Gamma \left( p \right)}},\tag{22}\]
where $\Gamma \left(  \cdot  \right)$ is Gamma function. When $p \in {{{\mathbb{Z}}}^ + }$, ${{{\mathbb{Z}}}^ + }$ is the set of positive integers, $\Gamma \left( p \right) = \left( {p - 1} \right)!$. The average value of ${R_p}$ is derived by
{\color{blue}
\[\begin{gathered}
\begin{aligned}
  {r_p} &= E\left( {{R_p}} \right) = \int_0^\infty  {r{f_{{R_p}}}\left( r \right)} dr \hfill \\
  &= \int_0^\infty  {r{e^{ - {\lambda _E}\pi {r^2}}}\frac{{2{{\left( {{\lambda _E}\pi {r^2}} \right)}^p}}}{{r\Gamma \left( p \right)}}} dr \hfill \\
  &= \frac{{\left( {2p - 1} \right)\left( {2p - 3} \right) \cdots 1}}{{\Gamma \left( p \right) \cdot {2^{p - 1}}}}\int_0^\infty  {{e^{ - {\lambda _E}\pi {r^2}}}} dr \hfill \\
  &\mathop  = \limits^{(b)} \frac{{\left( {2p - 1} \right)\left( {2p - 3} \right) \cdots 1}}{{\Gamma \left( p \right) \cdot {2^p} \cdot \sqrt {{\lambda _E}} }} \hfill \\
\end{aligned}
\end{gathered},\tag{23}\]
}\\
where $(b)$ results due to the Gaussian integral, i.e., $\int_0^\infty  {{e^{ - {\lambda _E}\pi {r^2}}}} dr = \frac{1}{{2\sqrt {{\lambda _E}} }}$.

In this paper, the relay routing algorithm with the minimum hop number is adopted for the transmission path between the requested EDC and the destination SBS. Therefore, the relay distance is configured as ${r_{mmW}}$. When the average transmission distance is ${r_p}$, the average hop number of ${r_p}$ is
\[{n_p} = \left\lceil {\frac{{{r_p}}}{{{r_{mmW}}}}} \right\rceil ,\tag{24}\]
where $\left\lceil  \cdot  \right\rceil $ is the rounding up operation of a number.

Considering the distributions of SBSs and EDCs, the number of SBSs in the coverage of a EDC is calculated by $1 + 1.28\frac{{{\lambda _S}}}{{{\lambda _E}}}$. Moreover, the probability that a SBS selected by the EDC for relaying the AR/VR data is ${p_1} = \frac{1}{{1 + 1.28\frac{{{\lambda _S}}}{{{\lambda _E}}}}}$\cite{R25}. When mmWave links are assumed to be line of sight (LoS) links in this paper, the wireless link fading is expressed by $L\left( {{\text{dB}}} \right) = 70 + 20{\log _{10}}\left( {{r_{mmW}}} \right) + \zeta $, $\zeta  \sim {\cal N}\left( {0,{\sigma ^2}} \right)$, where $\zeta$ is the shadow fading coefficient and $\sigma$ is the standard deviation of shadow fading over wireless channels in dB. The wireless transmission is successful over mmWave links only if ${P_S}\left( {dB} \right) - L\left( {dB} \right) - {N_0}{W_{mmW}}\left( {dB} \right) \geqslant {\theta _4}\left( {dB} \right)$ is satisfied. Therefore, successful probability of wireless transmission over mmWave links is \cite{R25}
\[\begin{gathered}
\begin{aligned}
  {p_2} &= P\left( {\zeta  \leqslant {P_S}\left( {dB} \right) - {\theta _4}\left( {dB} \right) - {N_0}{W_{mmW}}\left( {dB} \right)} \right.\left. { - 70 - 20{{\log }_{10}}\left( {{r_{mmW}}} \right)} \right) \hfill \\
  &= \frac{1}{2}\left( {1 + erf\left( {\frac{{f\left( {{r_{mmW}}} \right)}}{{\sqrt 2 \sigma }}} \right)} \right) \hfill \\
\end{aligned}
\end{gathered}.\tag{25}\]

When one packet is transmitted over the distance ${r_p}$ by the multi-hop relay method in this paper, the transmission delay is expressed by
\[\begin{gathered}
\begin{aligned}
  {D_p} &= {n_p} \cdot \frac{{{\tau _{mmW}}}}{{{p_1}{p_2}}} \hfill \\
  &= \left\lceil {\frac{{{r_p}}}{{{r_{mmW}}}}} \right\rceil {\tau _{mmW}}\left( {1 + 1.28\frac{{{\lambda _S}}}{{{\lambda _E}}}} \right)\frac{2}{{1 + erf\left( {\frac{{f\left( {{r_{mmW}}} \right)}}{{\sqrt 2 \sigma }}} \right)}} \hfill \\
  &\approx \frac{{{r_p}{\tau _{mmW}}}}{{{r_{mmW}}}}\left( {1 + 1.28\frac{{{\lambda _S}}}{{{\lambda _E}}}} \right)\frac{2}{{1 + erf\left( {\frac{{f\left( {{r_{mmW}}} \right)}}{{\sqrt 2 \sigma }}} \right)}} \hfill \\
\end{aligned}
\end{gathered}.\tag{26}\]

To reduce the effect of delay jitter caused by MCR scheme, a buffer is adopted at the destination SBS. When the SBS buffer size and the packet size are configured as $\Omega$ and $L$, the maximum tolerable buffer delay is $\left\lceil {\frac{\Omega}{L}} \right\rceil $ for a multi-hop relay route between the EDC and the destination SBS when the AR/VR application is run by users. When the transmission control protocol (TCP) is used for AR/VR data, the EDC can transmit the next packet only if the current packet is successfully accepted at the destination SBS. Hence, the total delay of one packet in the multi-hop relay route is $\left\lceil {\frac{\Omega}{L}} \right\rceil  \cdot {D_p}$. Based on the MCR scheme, the EDC $ED{C_p}$ only need to transmit the data of $\frac{{\frac{1}{{{r_p}}}}}{{\sum\limits_{i = 1}^B {\frac{1}{{{r_i}}}} }} \cdot \mathfrak{S}$. The system delay between the EDC $ED{C_p}$ and the destination SBS is expressed by
\[D_{_p}^w = \frac{{\frac{1}{{{r_p}}}}}{{\sum\limits_{i = 1}^B {\frac{1}{{{r_i}}}} }} \cdot \left\lceil {\frac{\Omega}{L}} \right\rceil  \cdot {D_p},\tag{27}\]
where ${r_i} = \frac{{\left( {2i - 1} \right)\left( {2i - 3} \right) \cdots 1}}{{\Gamma \left( i \right) \cdot {2^i} \cdot \sqrt {{\lambda _E}} }}$. Considering that $B$ EDCs are utilized for simultaneously transmission in the MCR scheme, the system delay of MCR scheme is derived by
\[\begin{gathered}
\begin{aligned}
  D_{DL}^{bh} &= \mathop {\max }\limits_{p = 1,2,...B} D_{_p}^w \hfill \\
  &= \mathop {\max }\limits_{p = 1,2,...B} \frac{{\left\lceil {\frac{\Omega}{L}} \right\rceil }}{{\sum\limits_{i = 1}^B {\frac{1}{{{r_i}}}} }} \cdot \frac{{2{\tau _{mmW}}\left( {1 + 1.28\frac{{{\lambda _S}}}{{{\lambda _E}}}} \right)}}{{{r_{mmW}}\left( {1 + erf\left( {\frac{{f\left( {{r_{mmW}}} \right)}}{{\sqrt 2 \sigma }}} \right)} \right)}} \hfill \\
  &\mathop  = \limits^{(c)} \frac{{\left\lceil {\frac{\Omega}{L}} \right\rceil }}{{\sum\limits_{i = 1}^B {\frac{1}{{{r_i}}}} }} \cdot \frac{{2{\tau _{mmW}}\left( {1 + 1.28\frac{{{\lambda _S}}}{{{\lambda _E}}}} \right)}}{{{r_{mmW}}\left( {1 + erf\left( {\frac{{f\left( {{r_{mmW}}} \right)}}{{\sqrt 2 \sigma }}} \right)} \right)}} \hfill \\
\end{aligned}
\end{gathered},\tag{28}\]
where $(c)$ is obtained under the condition that the system delay $D_p^w$ does not depend on the $p - th$ route based on the result of (27).

The MCR delay theorem is proved.

\textbf{Lemma 1}: When EDCs are deployed in a 5G small cell network, the AR/VR data stored at $B$ EDCs are simultaneously transmitted to a destination SBS by the MCR scheme. The lower and upper bounds of system delay in the MCR scheme are given by
\[\frac{{2.28\left\lceil {\frac{\Omega}{L}} \right\rceil {\tau _{mmW}}}}{{\pi R_{\max }^2\lambda _S^{\frac{3}{2}}{r_{mmW}}\left( {1 + erf\left( {\frac{{f\left( {{r_{mmW}}} \right)}}{{\sqrt 2 \sigma }}} \right)} \right)}} < D_{DL}^{bh} < \frac{{\left\lceil {\frac{\Omega}{L}} \right\rceil {\tau _{mmW}}\left( {1 + 1.28\frac{{{\lambda _S}}}{{{\lambda _M}}}} \right)}}{{\sqrt {{\lambda _M}} {r_{mmW}}\left( {1 + erf\left( {\frac{{f\left( {{r_{mmW}}} \right)}}{{\sqrt 2 \sigma }}} \right)} \right)}},\tag{29}\]
where ${R_{\max }}$ is the maximum distance between the EDC and the destination SBS.

\emph{\textbf{Proof:}} Based on the configuration of system model, densities of MBCs, SBSs and EDCs satisfy the following constraint: ${\lambda _M} < {\lambda _E} < {\lambda _S}$. The upper bound of system delay in the MCR scheme is derived by
\[\begin{gathered}
\begin{aligned}
  D_{DL}^{bh} &= \frac{{\left\lceil {\frac{\Omega}{L}} \right\rceil }}{{\sum\limits_{i = 1}^B {\frac{1}{{{r_i}}}} }} \cdot \frac{{2{\tau _{mmW}}\left( {1 + 1.28\frac{{{\lambda _S}}}{{{\lambda _E}}}} \right)}}{{{r_{mmW}}\left( {1 + erf\left( {\frac{{f\left( {{r_{mmW}}} \right)}}{{\sqrt 2 \sigma }}} \right)} \right)}} \hfill \\
  &\mathop  \leqslant \limits^{(d)} \frac{{\left\lceil {\frac{\Omega}{L}} \right\rceil }}{{\sum\limits_{i = 1}^1 {\frac{1}{{{r_i}}}} }} \cdot \frac{{2{\tau _{mmW}}\left( {1 + 1.28\frac{{{\lambda _S}}}{{{\lambda _E}}}} \right)}}{{{r_{mmW}}\left( {1 + erf\left( {\frac{{f\left( {{r_{mmW}}} \right)}}{{\sqrt 2 \sigma }}} \right)} \right)}} \hfill \\
  &= \frac{{\left\lceil {\frac{\Omega}{L}} \right\rceil {\tau _{mmW}}\left( {1 + 1.28\frac{{{\lambda _S}}}{{{\lambda _E}}}} \right)}}{{\sqrt {{\lambda _E}} {r_{mmW}}\left( {1 + erf\left( {\frac{{f\left( {{r_{mmW}}} \right)}}{{\sqrt 2 \sigma }}} \right)} \right)}} \hfill \\
  &\mathop  < \limits^{(e)} \frac{{\left\lceil {\frac{\Omega}{L}} \right\rceil {\tau _{mmW}}\left( {1 + 1.28\frac{{{\lambda _S}}}{{{\lambda _M}}}} \right)}}{{\sqrt {{\lambda _M}} {r_{mmW}}\left( {1 + erf\left( {\frac{{f\left( {{r_{mmW}}} \right)}}{{\sqrt 2 \sigma }}} \right)} \right)}} \hfill \\
\end{aligned}
\end{gathered},\tag{30}\]
where $(d)$ is obtained under the condition that the number of cooperative transmission EDCs is larger than one, i.e., $B > 1$, $({\text{e}})$ is obtained under the condition of ${\lambda _E} > {\lambda _M}$.

The average distance ${r_i}$ between the destination SBS and the EDC $ED{C_i}$ is derived by
{\color{blue}
\[\begin{gathered}
\begin{aligned}
  {r_i} &= \frac{{\left( {2i - 1} \right) \cdot \left( {2i - 3} \right) \cdots 1}}{{{2^i}\sqrt {{\lambda _E}}  \cdot \left( {i - 1} \right)!}} \hfill \\
  &= \frac{1}{{2\sqrt {{\lambda _E}} }} \cdot \frac{{2i - 1}}{{2i - 2}} \cdot \frac{{2i - 3}}{{2i - 4}} \cdots \frac{3}{2} \hfill \\
  &> \frac{1}{{2\sqrt {{\lambda _E}} }} \cdot \underbrace {1 \cdot 1 \cdots 1}_{i - 1} = \frac{1}{{2\sqrt {{\lambda _E}} }} \hfill \\
\end{aligned}
\end{gathered},\tag{31}\]
}\\
Based on the result of (31), the lower bound of system delay in the MCR scheme is derived by
\[\begin{gathered}
\begin{aligned}
  D_{DL}^{bh} &= \frac{{\left\lceil {\frac{\Omega}{L}} \right\rceil }}{{\sum\limits_{i = 1}^B {\frac{1}{{{r_i}}}} }} \cdot \frac{{2{\tau _{mmW}}\left( {1 + 1.28\frac{{{\lambda _S}}}{{{\lambda _E}}}} \right)}}{{{r_{mmW}}\left( {1 + erf\left( {\frac{{f\left( {{r_{mmW}}} \right)}}{{\sqrt 2 \sigma }}} \right)} \right)}} \hfill \\
  &> \frac{{\left\lceil {\frac{\Omega}{L}} \right\rceil }}{{\sum\limits_{i = 1}^B {2\sqrt {{\lambda _E}} } }} \cdot \frac{{2{\tau _{mmW}}\left( {1 + 1.28\frac{{{\lambda _S}}}{{{\lambda _E}}}} \right)}}{{{r_{mmW}}\left( {1 + erf\left( {\frac{{f\left( {{r_{mmW}}} \right)}}{{\sqrt 2 \sigma }}} \right)} \right)}} \hfill \\
  &= \frac{{\left\lceil {\frac{\Omega}{L}} \right\rceil }}{{B\sqrt {{\lambda _E}} }} \cdot \frac{{{\tau _{mmW}}\left( {1 + 1.28\frac{{{\lambda _S}}}{{{\lambda _E}}}} \right)}}{{{r_{mmW}}\left( {1 + erf\left( {\frac{{f\left( {{r_{mmW}}} \right)}}{{\sqrt 2 \sigma }}} \right)} \right)}} \hfill \\
  & \mathop  \geqslant \limits^{(f)} \frac{{\left\lceil {\frac{\Omega}{L}} \right\rceil }}{{{\lambda _E}\pi R_{\max }^2\sqrt {{\lambda _E}} }} \cdot \frac{{{\tau _{mmW}}\left( {1 + 1.28\frac{{{\lambda _S}}}{{{\lambda _E}}}} \right)}}{{{r_{mmW}}\left( {1 + erf\left( {\frac{{f\left( {{r_{mmW}}} \right)}}{{\sqrt 2 \sigma }}} \right)} \right)}} \hfill \\
  & \mathop  > \limits^{(g)} \frac{{2.28\left\lceil {\frac{\Omega}{L}} \right\rceil {\tau _{mmW}}}}{{\pi R_{\max }^2\lambda _S^{\frac{3}{2}}{r_{mmW}}\left( {1 + erf\left( {\frac{{f\left( {{r_{mmW}}} \right)}}{{\sqrt 2 \sigma }}} \right)} \right)}} \hfill \\
\end{aligned}
\end{gathered},\tag{32}\]
where $(f)$ is obtained under the condition of $B \leqslant {\lambda _E}\pi R_{\max }^2$, $(g)$ is obtained under the condition of ${\lambda _E} < {\lambda _S}$.

Therefore, the Lemma 1 is proved.

\section{Service Effective Energy Optimization}
\subsection{Service Effective Energy}
For the AR/VR applications, wireless transmissions are premised on the basis of QoS. In this paper, the QoS is defined by
\[QoS = 1\left\{ {D \leqslant {D_{\max }}} \right\},\tag{33}\]
where $1\left\{ {...} \right\}$ is an indicator function, which equals to 1 when the condition inside the bracket is satisfied and 0 otherwise; ${D_{\max }}$ is the maximum delay threshold for AR/VR applications. On the other hand, a massive amount of wireless traffic generated by AR/VR applications is transmitted in 5G small cell networks. Hence, the energy consumption is another important metric for evaluating the performance of the proposed MCR scheme. Considering the requirement of QoS in AR/VR applications, the service effective energy (SEE) is defined by
\[{E_{SEE}} = {E_{sys}} \cdot QoS,\tag{34}\]
where ${E_{sys}}$ is the system energy of MCR scheme.

Based on the system model in Fig.~\ref{fig1}, the system energy of MCR scheme includes the energy consumed at MBSs, SBSs and EDCs. Without loss of generality, the energy consumption of MBSs and SBSs is classified into the embodied energy, i.e., the energy consumed in the manufacturing process of infrastructure equipments from a life-cycle perspective, and the operation energy, i.e., the energy consumed for wireless traffic transmissions \cite{R28}. The energy consumption of EDCs is classified into the embodied energy, the operation energy and the storage energy, i.e., the energy consumed for video storage at EDCs. As a consequence, the system energy of MCR scheme is extended as
{\small{
\[\begin{gathered}
\begin{aligned}
  {E_{sys}} &= {\lambda _M}{E_{MBS}} + {\lambda _S}{E_{SBS}} + {\lambda _E}({E_{EDC}} + \Psi \cdot {E_{storage}}) \hfill \\
  &= {\lambda _M}\left( {P_{OP}^M \cdot T_{Lifetime}^M + E_{EM}^M} \right) + {\lambda _S}\left( {P_{OP}^S \cdot T_{Lifetime}^S + E_{EM}^S} \right) + {\lambda _E}\left( {P_{OP}^E \cdot T_{Lifetime}^E + E_{EM}^E} \right) \hfill \\
  &+ {\lambda _E}\Psi \cdot {E_{storage}} \hfill \\
  &= {\lambda _M}\left( {\left( {{a_M}{P_M} + {b_M}} \right) \cdot T_{Lifetime}^M + E_{EM}^M} \right) + {\lambda _S}\left( {\left( {{a_S}{P_S} + {b_S}} \right) \cdot T_{Lifetime}^S + E_{EM}^S} \right){\kern 1pt}  \hfill \\
  &+ {\lambda _E}\left( {\left( {{a_E}{P_E} + {b_E}} \right) \cdot T_{Lifetime}^E + E_{EM}^E} \right) + {\lambda _E}\Psi \cdot {E_{storage}} \hfill \\
\end{aligned}
\end{gathered},\normalsize{\tag{35}}\]
}}\\
where ${E_{MBS}}$, ${E_{SBS}}$ and ${E_{EDC}}$ are the energy consumption at MBS, SBS and EDC, respectively; ${E_{storage}}$ is the energy consumption of one video content stored at the EDC; $P_{OP}^M$, $P_{OP}^S$ and $P_{OP}^E$ are the operation power of MBS, SBS and EDC, respectively; $T_{Lifetime}^M$, $T_{Lifetime}^S$ and $T_{Lifetime}^E$ are the lifetime of MBS, SBS and EDC, respectively; $E_{EM}^M$, $E_{EM}^S$ and $E_{EM}^E$ are the embodied energy of MBS, SBS and EDC, respectively; ${a_M}$ and ${b_M}$ are the fixed coefficients of the operation power at MBSs; ${a_S}$ and ${b_S}$ are the fixed coefficient of operation power at SBSs; ${a_E}$ and ${b_E}$ are the fixed coefficients of operation power at EDCs.

\subsection{Algorithm Design}
Assumed that AR/VR video contents are stored in local EDCs. To save the energy consumption of MCR scheme, the optimal SEE problem is formulated by
\[\begin{gathered}
  \mathop {\min }\limits_{{\lambda _E},\Psi} {E_{SEE}} = {E_{sys}} \cdot QoS \hfill \\
  s.t.{\kern 1pt} {\kern 1pt} {\kern 1pt} \sum\limits_{k = 1}^\Psi {{x_{mk}}}  = \Psi,\forall ED{C_m} \in {\cal M} \hfill \\
  {\kern 1pt} {\kern 1pt} {\kern 1pt} {\kern 1pt} {\kern 1pt} {\kern 1pt} {\kern 1pt} {\kern 1pt} {\kern 1pt} {\kern 1pt} {\kern 1pt} {\kern 1pt} {\kern 1pt} {\kern 1pt} {\kern 1pt} {r_i} \leqslant {R_{\max }} \hfill \\
  {\kern 1pt} {\kern 1pt} {\kern 1pt} {\kern 1pt} {\kern 1pt} {\kern 1pt} {\kern 1pt} {\kern 1pt} {\kern 1pt} {\kern 1pt} {\kern 1pt} {\kern 1pt} {\kern 1pt} {\kern 1pt} {\kern 1pt} {P_M} > {P_S} > {P_U} \hfill \\
\end{gathered},\tag{36}\]
where the minimum SEE is solved by finding the optimal density of EDCs ${\lambda _E}$ and the optimal number of video contents $\Psi$ at EDCs. Considering the popularity distribution of video contents, the total number of video contents $\Psi$ is expressed by $\sum\limits_{k = 1}^\Psi {{x_{mk}}}  = \Psi, \forall ED{C_m} \in {\cal M}$. To avoid the difficulty caused by an infinite distance between the EDC and the destination SBS on solving the optimization problem, the maximum distance between the EDC and the destination SBS is constrained within a maximum threshold ${R_{\max }}$ for the optimal SEE. Considering functions of MBSs, SBSs and users, the wireless transmission powers of MBSs, SBSs and users are constrained by ${P_M} > {P_S} > {P_U}$.

To solve the optimal SEE problem in (36), a two-step solution is proposed in this paper. In Step 1, the required system delay is solved for the AR/VR MCR scheme. In Step 2, the SEE is optimized for the AR/VR MCR scheme.

Step 1: Based on (36), the required system delay is formulated by
\[\begin{gathered}
  \mathop {\max }\limits_{{\lambda _E},A} QoS \hfill \\
  s.t.{\kern 1pt} {\kern 1pt} {\kern 1pt} \sum\limits_{k = 1}^\Psi {{x_{mk}}}  = \Psi,\forall ED{C_m} \in {\cal M} \hfill \\
  {\kern 1pt} {\kern 1pt} {\kern 1pt} {\kern 1pt} {\kern 1pt} {\kern 1pt} {\kern 1pt} {\kern 1pt} {\kern 1pt} {\kern 1pt} {\kern 1pt} {\kern 1pt} {\kern 1pt} {\kern 1pt} {r_i} \leqslant {R_{\max }} \hfill \\
  {\kern 1pt} {\kern 1pt} {\kern 1pt} {\kern 1pt} {\kern 1pt} {\kern 1pt} {\kern 1pt} {\kern 1pt} {\kern 1pt} {\kern 1pt} {\kern 1pt} {\kern 1pt} {\kern 1pt} {\kern 1pt} {\kern 1pt} {P_M} > {P_S} > {P_U} \hfill \\
\end{gathered}.\tag{37}\]
Based on (3),  the delays $D_{UL}^{req}$, $D_{DL}^{deli}$ and $D_{DL}^{as}$ are independent of the density of EDCs ${\lambda _E}$ and the number of video contents $\Psi$. Therefore, in this optimal algorithm the maximum delay threshold of AR/VR applications is replaced by a variable $D{'_{\max }} = {D_{\max }} - D_{UL}^{req} - D_{DL}^{deli} - D_{DL}^{as}$. The condition inside of QoS indication function is expressed by
\[D_{DL}^{bh} + {D^{fiber}} \cdot \left( {1 - {P_{in - EDC}}} \right) \leqslant D{'_{\max }}.\tag{38}\]
Based on (20) and (21), the backhaul delay $D_{DL}^{bh}$ decreases with the increase of the EDC density ${\lambda _E}$ and the fiber link delay ${D^{fl}} = {D^{fiber}} \cdot \left( {1 - {P_{in - EDC}}} \right)$ decreases with the increases in the number of video contents $\Psi$. Considering two conditions ${\lambda _E} > 0$ and $1 \leqslant \Psi \leqslant \left| {\cal K} \right|,\Psi \in {{{\mathbb{Z}}}^ + }$, the critical value $\lambda _E^\Psi$ with the given value of $\Psi$ is obtained by traversing all available values in the set of $\Psi = \left\{ {1,2,...,\left| {\cal K} \right|} \right\}$. When a value of $\Psi = 1,...,\left| {\cal K} \right|$ is substituted into (38), a corresponding value of ${\lambda _E} \geqslant \lambda _E^\Psi$ is obtained. Therefore, $\lambda _E^\Psi$ is the critical value for the available value of $\Psi$.  Moreover, the available value pair of ${\lambda _E}$ and $\Psi$ is denoted by ${\cal C}\left( {\Psi,{\lambda _E}} \right) = \left\{ {(\Psi,\lambda _E^\Psi)|\Psi = 1,...,\left| {\cal K} \right|} \right\}$.

Step 2: Based on the result of Step 1, the minimum system energy of MCR scheme is formulated by
\[\begin{gathered}
  \mathop {\min }\limits_{{\lambda _E},S} {E_{sys}} \hfill \\
  s.t.{\kern 1pt} {\kern 1pt} {\kern 1pt} \left( {\Psi,{\lambda _E}} \right) \in {\cal C}\left( {\Psi,{\lambda _E}} \right) \hfill \\
\end{gathered},\tag{39}\]
where the available value pairs $\Psi$ and $\lambda _E^\Psi$ are substituted into (35), the optimal SEE is solved by obtaining the minimum system energy of MCR scheme. The detailed SEE optiMization (SEEM) algorithm is shown in Algorithm 1.
\begin{algorithm*}
\setcounter{algorithm}{0}
\renewcommand{\algorithmicrequire}{\textbf{Input:}}
\renewcommand\algorithmicensure {\textbf{Output:} }

\begin{algorithmic}
\caption{Service Effective Energy optiMization (SEEM) Algorithm}

\REQUIRE ${D^{fiber}}$, $D{'_{\max }}$ , and relevant parameters about $D_{DL}^{bh},{{P_{in - EDC}}}$\\
\ENSURE The minimum value of ${E_{sys}}$, i.e., ${E_{sys\_\min }}$, the value pair {\color{blue} ${vp_x}$}, which denotes an available value pair {\color{blue}$\left( {\Psi,\lambda _E\left( \Psi \right)} \right)$} to get ${E_{sys\_\min }}$\\
\textbf{Initialization:} $i = 0;\Phi  = \phi ;$
\begin{enumerate}
\item \textbf{for} {$\Psi = 1:\left| {{\mathcal{K}}} \right|$} \textbf{do} \\
\quad \textbf{if} {${D^{fiber}} \cdot \left( {1 - {P_{in - EDC}}} \right) > D{'_{\max }}$} \textbf{then}\\
{\color{blue} \[\lambda _E\left( \Psi \right) \leftarrow 0;\] }
\quad \textbf{else}\\
\quad \quad Compute equation $D_{DL}^{bh} + {D^{fiber}} \cdot \left( {1 - {P_{in - EDC}}} \right) = D{'_{\max }}$ with value $\Psi$ to obtain the critical value ${\lambda _{cr}}$;
{\color{blue} \[\lambda _E\left( \Psi \right) \leftarrow {\lambda _{cr}};\] }
\quad \textbf{end if}\\
\textbf{end for}\\
\item \textbf{for} {$\Psi = 1:\left| {{\mathcal{K}}} \right|$} \textbf{do} \\
\quad \textbf{if} {\color{blue} {$\lambda _E\left( \Psi \right) =  = 0$} } \textbf{then}\\
\quad\quad Continue;\\
\quad \textbf{else}\\
\[i \leftarrow i + 1;{\color{blue} {vp_i} \leftarrow \left( {\Psi,\lambda _E\left( \Psi \right)} \right); }\]
\quad\quad Put {\color{blue}${vp_i}$} into the set $\Phi $;\\
\quad \textbf{end if}\\
\textbf{end for}\\
{\color{blue}
\[ {vp_x}  \leftarrow \arg \mathop {\min }\limits_{{vp_i} \in \Phi } {E_{sys}}\left( {{vp_i}} \right);\]
\[{E_{sys\_\min }} \leftarrow \mathop {\min }\limits_{{vp_i} \in \Phi } {E_{sys}}\left( {{vp_i}} \right);\]
}
\end{enumerate}
\end{algorithmic}
\end{algorithm*}

\section{Simulation Results and Performance Analysis}
Based on the proposed system delay model of the MCR scheme, the effect of various system parameters on the system delay of the MCR scheme will be analyzed and compared by numerical simulations in this section. In what follows, the default values of system model are illustrated in Table~\ref{tab1}. Moreover, the performance of SEEO algorithm is simulated and analyzed in this section.

\begin{figure}[!t]
\centerline{\includegraphics[width=9cm, draft=false]{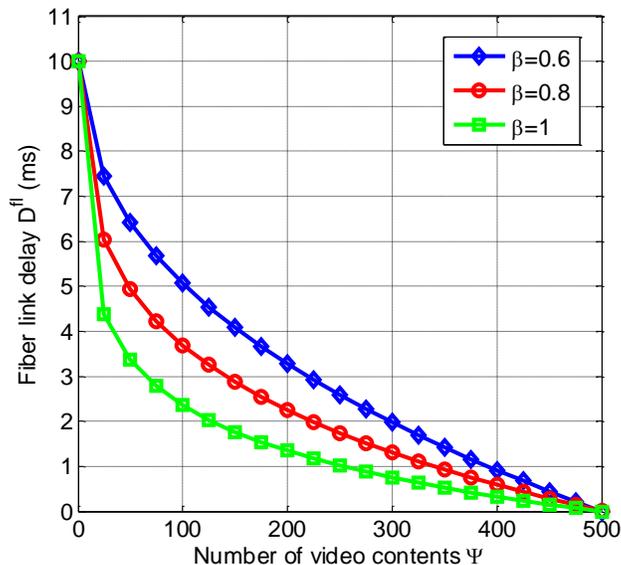}}
\caption{Fiber link delay with respect to the number of video contents under different skewness parameters of popularity distributions.}
\label{fig2}
\end{figure}

Fig.~\ref{fig2} shows the fiber link delay with respect to the number of video contents considering different skewness parameters of popularity distributions. When the skewness parameter of popularity distribution is fixed, the fiber link delay decreases with the increase of the number of video content at EDCs. When the number of video content is fixed, the fiber link delay decreases with the increase of the skewness parameter of popularity distribution.

\begin{figure}[!t]
\centerline{\includegraphics[width=9cm, draft=false]{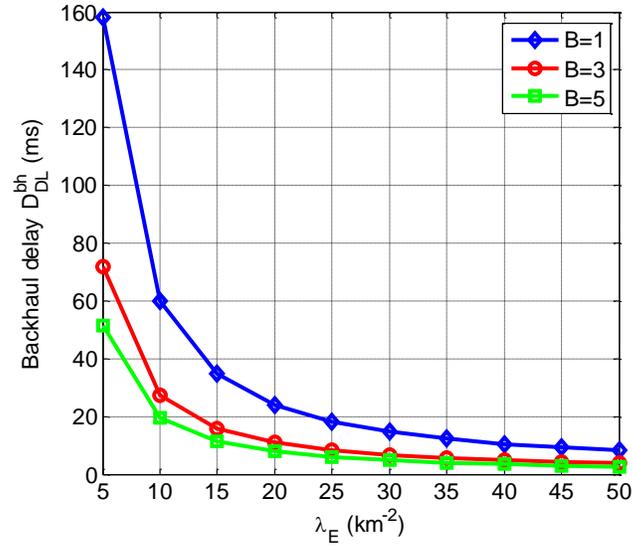}}
\caption{Backhaul delay with respect to the density of EDCs under different numbers of cooperative EDCs.}
\label{fig3}
\end{figure}

Fig.~\ref{fig3} depicts the backhaul delay with respect to the density of EDCs considering different number of cooperative EDCs. When the number of cooperative EDCs is fixed, the backhaul delay decreases with the increase of the density of EDCs. When the density of EDCs is fixed, the backhaul delay decreases with the increase of the number of cooperative EDCs.

\begin{figure}[!t]
\centerline{\includegraphics[width=9cm, draft=false]{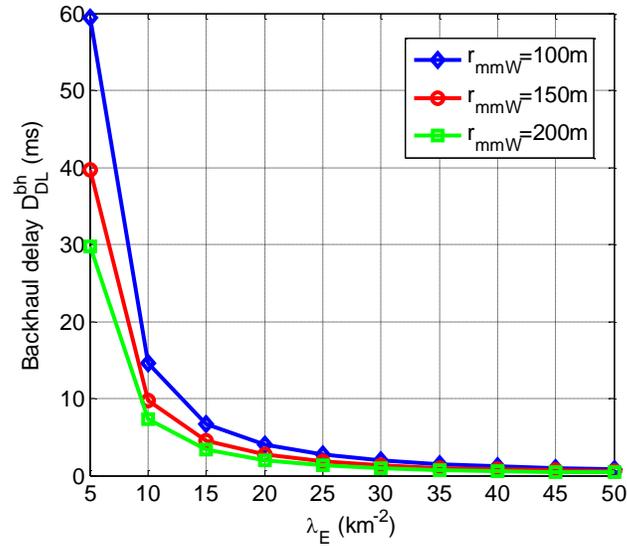}}
\caption{Backhaul delay with respect to the density of EDCs under different maximum transmission distances of SBSs and EDCs.}
\label{fig4}
\end{figure}

Fig.~\ref{fig4} illustrates the backhaul delay with respect to the density of EDCs under different maximum transmission distances of SBSs and EDCs. When the density of EDCs is fixed, the backhaul delay decreases with the increase of the maximum transmission distances of SBSs and EDCs.

\begin{figure*}[!t]
\centerline{\includegraphics[width=17cm, draft=false]{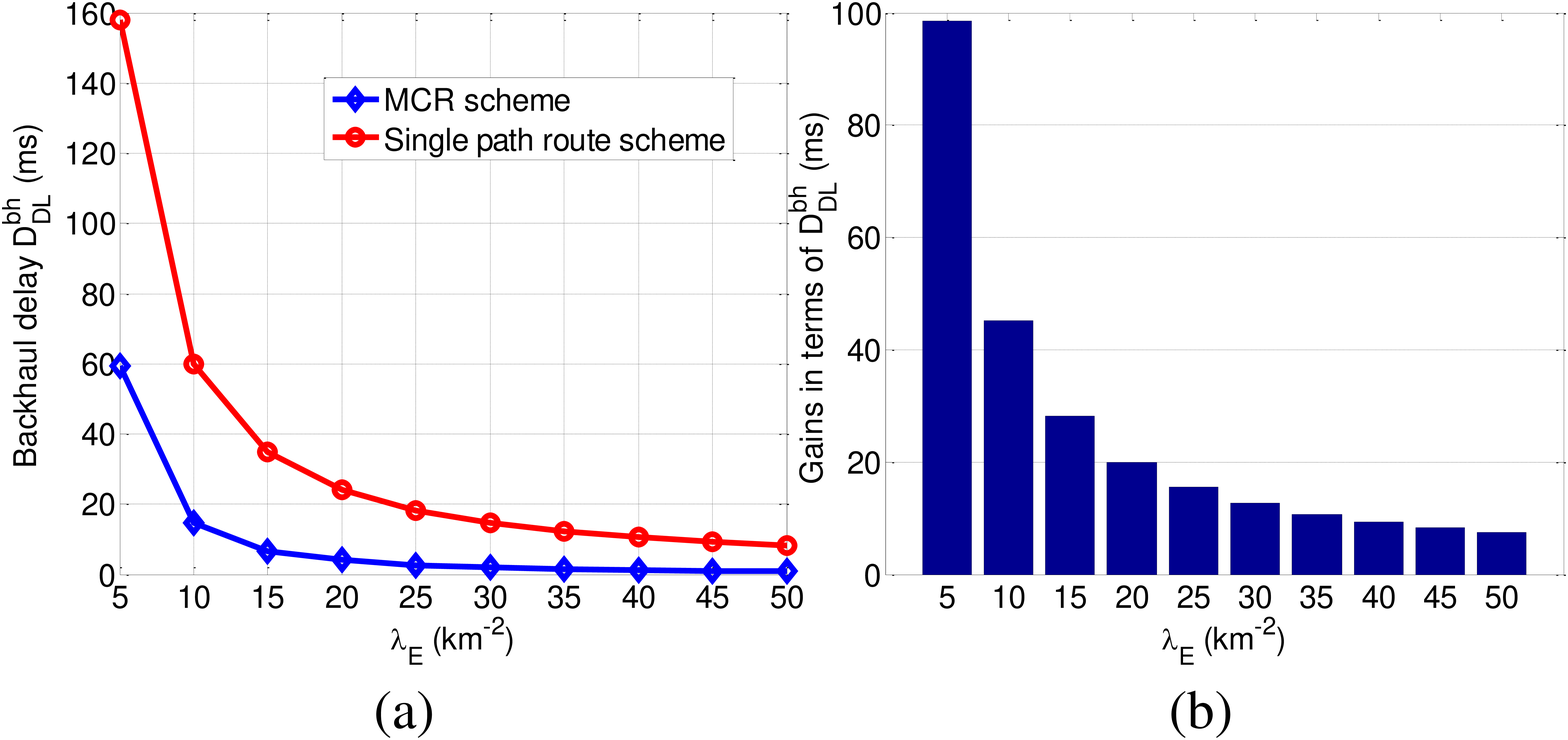}}
\caption{Backhaul delay with respect to the density of EDCs under the MCR scheme and the single path route scheme.}
\label{fig5}
\end{figure*}

Fig.~\ref{fig5} compares the backhaul delay with respect to the density of EDCs under the MCR scheme and the single path route scheme. Fig.~\ref{fig5}(a) shows that the backhaul delay of MCR scheme is less than that of the single path route scheme. Fig.~\ref{fig5}(b) describes the {\color{blue}gains in terms of} backhaul delay achieved by the MCR scheme over the single path route scheme. Results in Fig.~\ref{fig5}(b) indicate that the {\color{blue}gains in terms of} backhaul delay decrease with the increase of the density of EDCs.

\begin{figure}[!t]
\centerline{\includegraphics[width=9cm, draft=false]{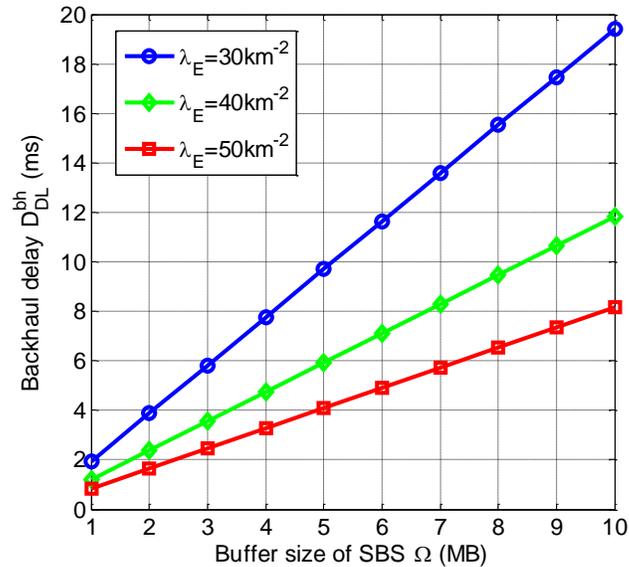}}
\caption{Impact of buffer size of SBSs on the backhaul delay with different densities of EDCs.}
\label{fig6}
\end{figure}

In Fig.~\ref{fig6}, the impact of buffer size of SBSs on the backhaul delay with different densities of EDCs is investigated. When the density of EDCs is fixed, the backhaul delay increases with the increase of the buffer size of SBSs.

\begin{figure}[!t]
\centerline{\includegraphics[width=9cm, draft=false]{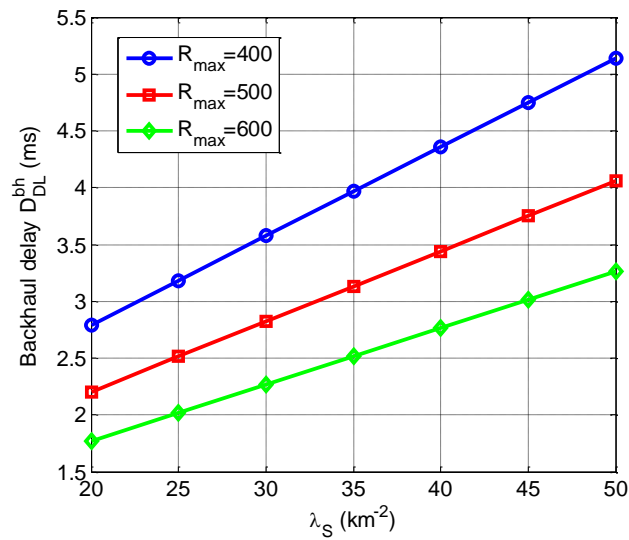}}
\caption{Backhaul delay with respect to the density of SBSs under different maximum distances between the EDC and the destination SBS.}
\label{fig7}
\end{figure}

Fig.~\ref{fig7} presents the backhaul delay with respect to the density of SBSs under different maximum distances between the EDC and the destination SBS. When the maximum distance between the EDC and the destination SBS is fixed, the backhaul delay increases with the density of SBSs. When the density of SBSs is fixed, the backhaul delay decreases with the increase of the maximum distance between the EDC and the destination SBS.

\begin{figure}[!t]
\centerline{\includegraphics[width=9cm, draft=false]{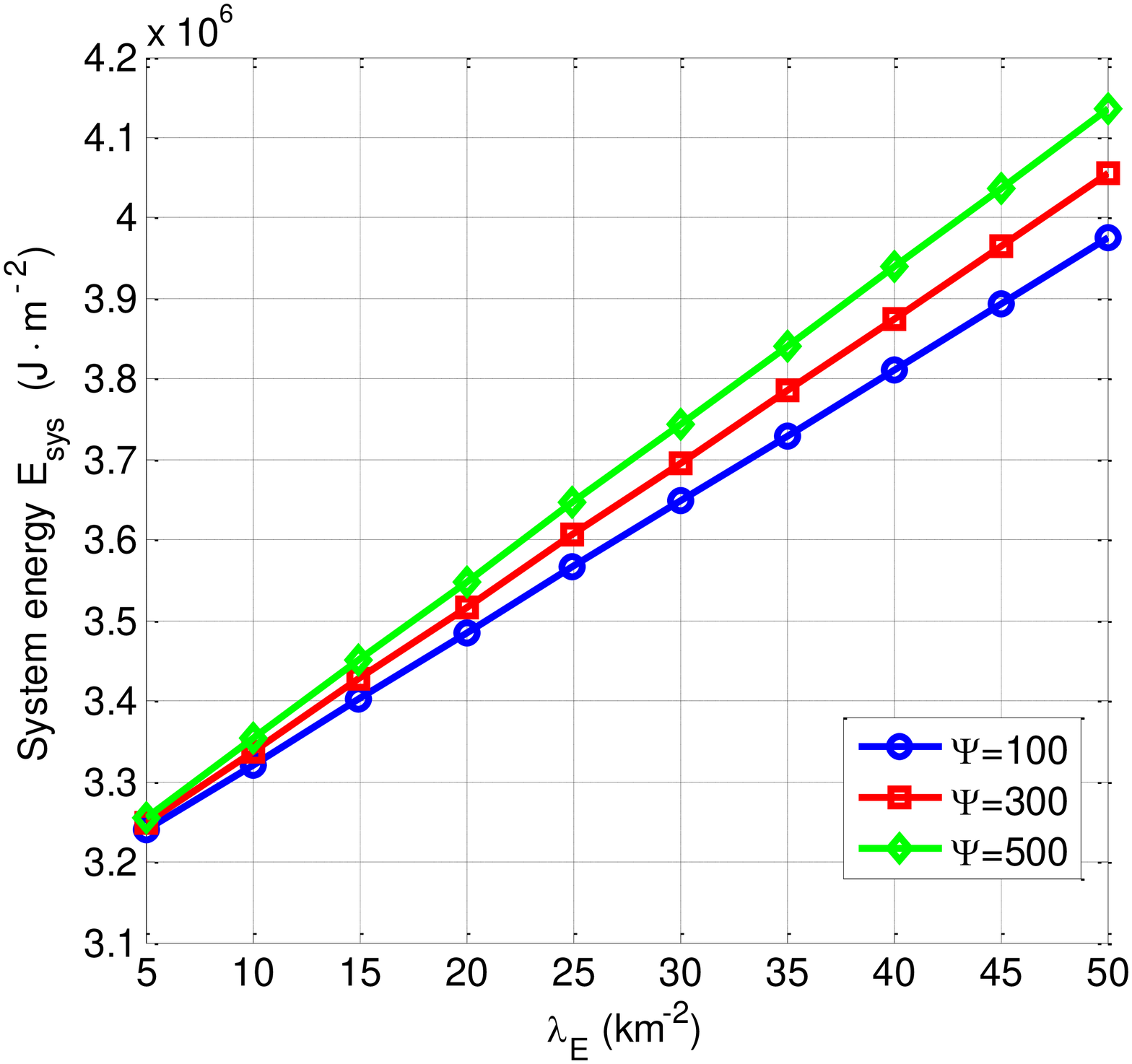}}
\caption{System energy with respect to the density of EDCs under different numbers of video contents stored at EDCs.}
\label{fig8}
\end{figure}

Fig.~\ref{fig8} shows the system energy with respect to the density of EDCs under different numbers of video contents stored at EDCs. When the number of video contents is fixed, the system energy increases with the density of EDCs. When the density of EDCs is fixed, the system energy increases with the increase of the number of video contents stored at EDCs.

\begin{figure*}[!t]
\centerline{\includegraphics[width=17cm, draft=false]{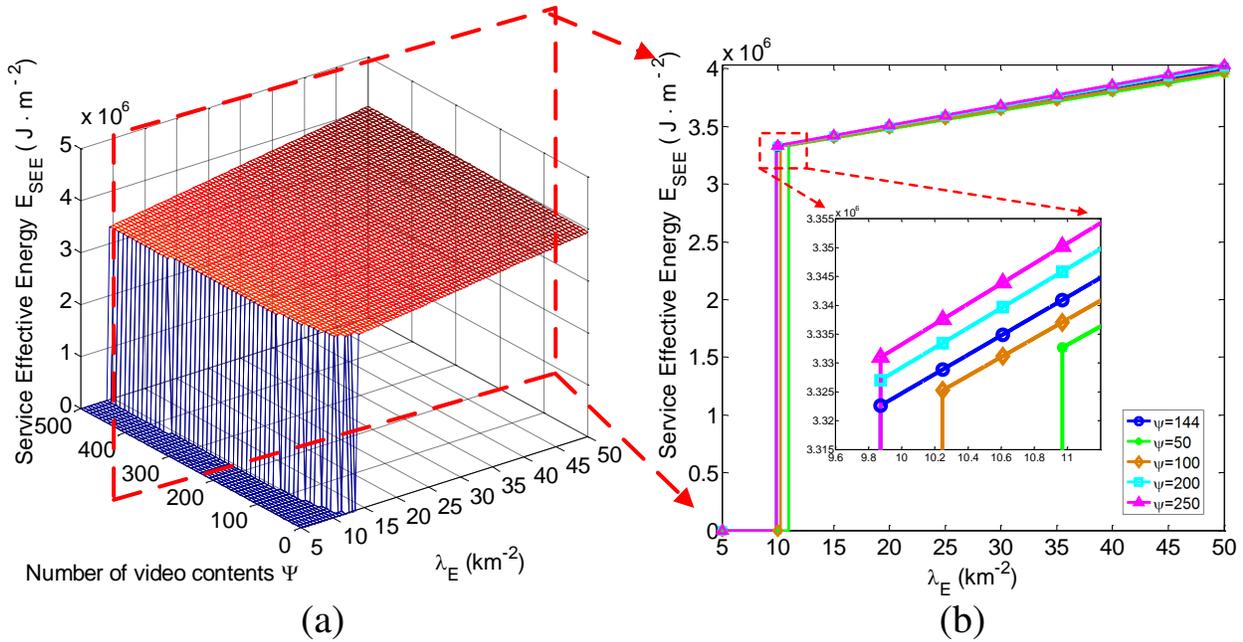}}
\caption{SEE with respect to the density of EDCs under different numbers of video contents stored at EDCs.}
\label{fig9}
\end{figure*}

Fig.~\ref{fig9} depicts the SEE with respect to the density of EDCs under different numbers of video contents stored at EDCs. Based on the SEEO algorithm, the optimal solution of $\Psi$ and $\lambda _E^\Psi$ are solved by $\Psi = 144$ and $\lambda _E^\Psi = 9.873$ ${\text{k}}{{\text{m}}^{{\text{-2}}}}$. Fig.~\ref{fig9}(a) is a three-dimension figure describing the relationship among the SEE, the density of EDCs and the number of video contents stored at EDCs. Fig.~\ref{fig9}(b) provides a two-dimension view of Fig. 9 (a),  for a better illustration. Based on the results in Fig. 9, the minimum SEE is achieved when the number of video contents and the density of EDCs are configured as 144 and 9.873 ${\text{k}}{{\text{m}}^{{\text{-2}}}}$, respectively.

\begin{figure}[!t]
\centerline{\includegraphics[width=9cm, draft=false]{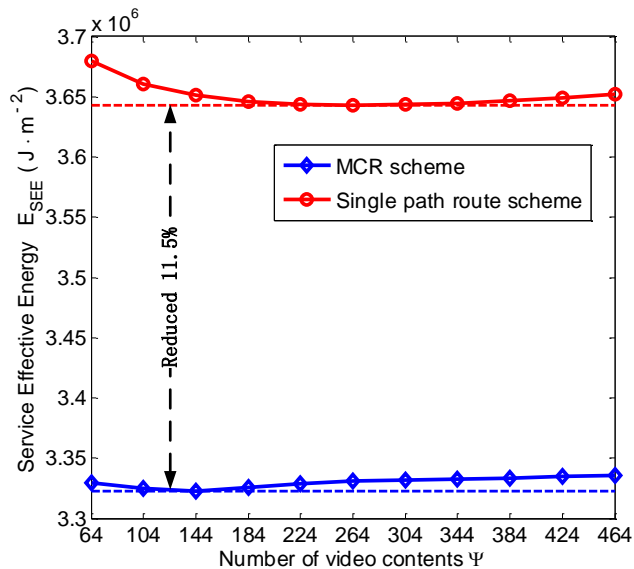}}
\caption{SEE with respect to the number of video contents under the MCR scheme and the single path route scheme.}
\label{fig10}
\end{figure}

Fig.~\ref{fig10} compares the SEE with respect to the number of video contents under the MCR scheme and the single path route scheme. Based on the curves in Fig.~\ref{fig10}, the SEE of MCR scheme is always less than that of single path route scheme in 5G small cell networks. When the number of video contents stored at EDCs is configured as 144, the SEE of MCR scheme achieves the minimum, i.e., $3.3226 \times {10^6}{\kern 1pt} {\kern 1pt} {\text{J}} \cdot {{\text{m}}^{{\text{-2}}}}$. When the number of video contents stored at EDCs is configured as 264, the SEE of single path route scheme achieves the minimum, i.e., $3.6432 \times {10^6}{\kern 1pt} {\kern 1pt} {\text{J}} \cdot {{\text{m}}^{{\text{-2}}}}$. Compared with the SEE minimum of single path route scheme, the SEE minimum of MCR scheme is reduced by 11.5\%.

\section{Conclusions}
The requirement of lower latency and massive data transmission for AR/VR applications imposes a great challenge on future wireless networks. In this paper a solution based on the SDN architecture is proposed for facilitating the AR/VR service provisionings in 5G small cell networks. To meet the requirements of lower delay and massive data transmission in AR/VR applications, a MCR scheme is proposed with an objective of achieving more efficient AR/VR wireless transmissions in 5G small cell networks, in which, a theorem on the delay of MCR scheme is proposed. Furthermore, both the lower and upper bounds of the delay in the MCR scheme are derived. {\color{blue}Since VR technologies enable the user to interact with the virtual world, VR technologies are more sensitive to the latency compared with AR technologies. The SEEM algorithm is designed to minimize the network energy consumption while guaranteeing that the delay is less than a given threshold by adopting the MCR scheme. Therefore, the SEEM algorithm is relatively more suitable for VR applications than for AR applications. }Simulation results indicate performance gains on both the delay and the SEE of the proposed MCR scheme compared with that of the conventional single path routing scheme in future 5G small cell networks.


\begin{thebibliography}{1}

\bibitem{R1}
Qualcomm, "Whitepaper: Making immersive virtual reality possible in mobile", Apr. 2016[Online]. Available: https://www.qualcomm.com/media/documents/files/making-immersive-virtual-reality-possible-in-mobile.pdf
\bibitem{R2}
Huawei, "Whitepaper on the VR-oriented bearer network requirement", Sep. 2016[Online]. Available: http://www-file.huawei.com/$\sim$/media/CORPORATE/PDF/white$\%20$paper/whitepaper-on-the-vr-oriented-bearer-network-requirement-en.pdf
\bibitem{R3}
Huawei, "Whitepaper: 5G opening up new business opportunities", Aug. 2016[Online]. Available: http://www.huawei.com/minisite/hwmbbf16/insights/5g-opening-up-new-business-opportunities-en.pdf
\bibitem{R4}
N. Bhushan, J. Li, D. Malladi and R. Gilmore, "Network densification: The dominant theme for wireless evolution into 5G," \emph{IEEE Commun. Mag.}, vol. 52, no. 2, pp. 82--89, Feb. 2014.
\bibitem{R5}
E. G. Larsson, O. Edfors, F. Tufvesson and T. L. Marzetta, "Massive MIMO for next generation wireless systems," \emph{IEEE Commun. Mag.}, vol. 52, no. 2, pp. 186--195, Feb. 2014.
\bibitem{R6}
T. Bai and R. W. Heath, "Coverage and rate analysis for millimeter-wave cellular networks," \emph{IEEE Trans. Wireless Commun.}, vol. 14, no. 2, pp. 1100--1114, Feb. 2015.
\bibitem{R7}
T. S. Rappaport \emph{et al.}, "Millimeter wave mobile communications for 5G cellular: It will work!," \emph{IEEE Access}, vol. 1, pp. 335--349, 2013.
\bibitem{R8}
C. Walravens and B. Gaidioz, "Breaking up gigabit ethernet's VoIP bottlenecks," \emph{IEEE Potentials}, vol. 27, no. 1, pp. 12--17, Jan.-Feb. 2008.
\bibitem{R9}
N. Laoutaris, G. Smaragdakis, R. Stanojevic, P. Rodriguez, and R. Sundaram, "Delay-tolerant bulk data transfers on the Internet," \emph{IEEE/ACM Trans. Netw.}, vol. 21, no. 6, pp. 1852--1865, Dec. 2013.
\bibitem{R10}
M. Villari, M. Fazio, S. Dustdar, O. Rana and R. Ranjan, "Osmotic computing: A new paradigm for edge/cloud Integration," \emph{IEEE Cloud Computing}, vol. 3, no. 6, pp. 76--83, Nov.-Dec. 2016.
{\color{blue}
\bibitem{N1}
W. Xiang, G. Wang, M. Pickering and Y. Zhang, "Big video data for light-field-based 3D telemedicine," \emph{IEEE Netw.}, vol. 30, no. 3, pp. 30--38, May-Jun. 2016.
\bibitem{N2}
G. Wang, W. Xiang, M. Pickering and C. W. Chen, "Light field multi-view video coding with two-directional parallel inter-view prediction," \emph{IEEE Trans. on Image Process.}, vol. 25, no. 11, pp. 5104--5117, Nov. 2016.
}
\bibitem{R11}
V. Bobrovs, S. Spolitis, G. Ivanovs, "Latency causes and reduction in optical metro networks," \emph{Proceeding of SPIE}, vol. 9008, pp. 117--124, 2014.
\bibitem{R12}
J.-C. Tsai, "Rate control for low-delay video coding using a dynamic rate table," \emph{IEEE Trans. Circuits Syst. Video Technol.}, vol. 15, no. 1, pp. 133--137, Jan. 2005.
\bibitem{R13}
R. Razavi, M. Fleury, and M. Ghanbari, "Low-delay video control in a personal area network for augmented reality," \emph{IET Image Process.}, vol. 2, no. 3, pp. 150--162, Jun. 2008.
\bibitem{R14}
A. D. Hartl, C. Arth, J. Grubert and D. Schmalstieg, "Efficient verification of holograms using mobile augmented reality," \emph{IEEE Trans. Visu. Computer Grap.}, vol. 22, no. 7, pp. 1843--1851, July 2016.
\bibitem{R15}
Y. A. Sekhavat, "Privacy preserving cloth try-on using mobile augmented reality," \emph{IEEE Trans. Multimedia}, to be published, doi: 10.1109/TMM.2016.2639380.
\bibitem{R16}
X. Li, N. Jolani, T. T. Dao and H. Jimison, "Serenity: A low-cost and patient-guided mobile virtual Reality Intervention for Cancer Coping," in \emph{Proc. IEEE ICHI 2016}, Chicago, IL, pp. 504--510, 2016.
\bibitem{R17}
J. E. Mu${\tilde {\text n}}$oz, T. Paulino, H. Vasanth and K. Baras, "PhysioVR: A novel mobile virtual reality framework for physiological computing," in \emph{Proc. IEEE Healthcom 2016}, Munich, pp. 1--6, 2016.
\bibitem{R18}
S. W. Choi, M. W. Seo and S. J. Kang, "Prediction-based latency compensation technique for head mounted display," in \emph{Proc. IEEE ISOCC 2016}, Jeju, South Korea, pp. 9--10, 2016.
\bibitem{R19}
T. Langlotz, M. Cook and H. Regenbrecht, "Real-time radiometric compensation for optical see-through head-mounted displays," \emph{IEEE Trans. Visu. Computer Grap.}, vol. 22, no. 11, pp. 2385--2394, Nov. 2016.
\bibitem{R20}
J. Lee, S. Park, I. Hong and H. J. Yoo, "A 3.13nJ/sample energy-efficient speech extraction processor for robust speech recognition in mobile head-mounted display systems," in \emph{Proc. IEEE ISCAS 2015}, Lisbon, pp. 1790--1793, 2015.
\bibitem{R21}
J. Li, J. Sun, Y. Qian, F. Shu, M. Xiao and W. Xiang, "A commercial video-caching system for small-cell cellular networks using game theory," \emph{IEEE Access}, vol. 4, pp. 7519--7531, 2016.
\bibitem{R22}
X. Ge, H. Cheng, G. Mao, Y. Yang and S. Tu, "Vehicular communications for 5G cooperative small-cell networks," \emph{IEEE Trans. Veh. Technol.}, vol. 65, no. 10, pp. 7882--7894, Oct. 2016.
\bibitem{R23}
D. Hu, J. Wu and P. Fan, "Minimizing end-to-end delays in linear multihop networks," \emph{IEEE Trans. Veh. Technol.}, vol. 65, no. 8, pp. 6487--6496, Aug. 2016.
\bibitem{R24}
S. Singh, M. N. Kulkarni, A. Ghosh and J. G. Andrews, "Tractable model for rate in self-backhauled millimeter wave cellular networks," \emph{IEEE J. Sel. Areas Commun.}, vol. 33, no. 10, pp. 2196--2211, Oct. 2015.
\bibitem{R25}
G. Zhang, T. Q. S. Quek, M. Kountouris, A. Huang and H. Shan, "Fundamentals of heterogeneous backhaul design-analysis and optimization," \emph{IEEE Trans. Commun.}, vol. 64, no. 2, pp. 876--889, Feb. 2016.
\bibitem{R26}
J. Liu, H. Nishiyama, N. Kato and J. Guo, "On the outage probability of device-to-device-communication-enabled multichannel cellular networks: An RSS-threshold-based perspective," \emph{IEEE J. Sel. Areas Commun.}, vol. 34, no. 1, pp. 163--175, Jan. 2016.
\bibitem{R27}
I. Banerjee, I. Roy, A. R. Choudhury, B. D. Sharma and T. Samanta, "Shortest path based geographical routing algorithm in wireless sensor network," in \emph{Proc. IEEE CODIS 2012}, Kolkata, pp. 262--265, Dec. 2012.
\bibitem{R28}
X. Ge, B. Yang, J. Ye, G. Mao, C.-X. Wang and T. Han, "Spatial spectrum and energy efficiency of random cellular networks," \emph{IEEE Trans. Commun.}, vol. 63, no. 3, pp. 1019--1030, Mar. 2015.


\end{thebibliography}
\end{document}